\documentclass[apj,twocolumn,twocolappendix,numberedappendix]{openjournal}
\usepackage{newtxtext,newtxmath,enumitem,multirow}
\usepackage[utf8]{inputenc}
\usepackage[T1]{fontenc}
\usepackage[breaklinks,colorlinks,allcolors=blue,citecolor=blue,urlcolor=blue]{hyperref}
\usepackage{orcidlink}
\pdfoutput=1

\def\GRUMPY{\texttt{GRUMPY\,}}

\def\zrei{z_{\rm rei}}
\def\M200c{M_{\rm 200c}}
\def\Muv{M_{1500}}
\def\Luv{L_{1500}}

\def\Ls{L_{*}}
\def\Lt{L_{t}}
\def\phis{\phi_{*}}

\def\ssfr10{\rm sSFR_{10}}
\def\nion{n_{\rm ion}}
\def\dnion{\dot{n}_{\rm ion}}
\def\dNion{\dot{N}_{\rm ion}}
\def\fesc{f_{\rm esc}}
\def\xiion{\xi_{\rm ion}}
\def\aa{\rm \mathring{A}}

\shorttitle{Modeling reionization history of the Universe}
\shortauthors{Wu, Kravtsov \& Katz}

\begin{document}
\title[Dwarfs and fesc on reionization]{Effect of ionizing photon escape fraction in faint galaxies\\ on modeling reionization history of the universe\vspace{-1.5cm}}

\author{Zewei Wu\,\orcidlink{0000-0002-7944-2543}$^{1,2\star}$}
\author{Andrey Kravtsov\,\orcidlink{0000-0003-4307-634X}$^{1,3,4,\dagger}$\vspace{2mm}}
\author{Harley Katz\,\orcidlink{0000-0003-1561-3814}$^{1,3}$\vspace{2mm}}

\affiliation{$^{1}$Department of Astronomy  \& Astrophysics, The University of Chicago, Chicago, IL 60637 USA}
\affiliation{$^{2}$Department of Astronomy  \& Astrophysics, University of California, San Diego, La Jolla, CA 92093 USA}
\affiliation{$^{3}$Kavli Institute for Cosmological Physics, The University of Chicago, Chicago, IL 60637 USA}
\affiliation{$^{4}$Enrico Fermi Institute, The University of Chicago, Chicago, IL 60637 USA}
\thanks{
$^\star$\href{mailto:jaw064@ucsd.edu}{jaw064@ucsd.edu}, $^\dagger$\href{mailto:kravtsov@uchicago.edu}{kravtsov@uchicago.edu}
}

\begin{abstract}
We present model calculations of the reionization history of hydrogen using star formation histories, computed with a galaxy formation model which reproduces properties of local dwarf galaxies and UV luminosity functions of galaxies at $z=5-16$. We use the ionizing photon density functions predicted by the model along with different models for the escape fraction of ionizing photons, $f_{\rm esc}$, to study the effects of ionizing photons from faint galaxies and different assumptions about $f_{\rm esc}$ on the evolution of hydrogen ionized fraction with redshift, $Q_{\rm HII}(z)$. We show that accounting for the contribution of faint galaxies with UV luminosities $M_{1500}>-13$, and with a {\it constant} ionizing photon escape fraction of $f_{\rm esc}=0.1$ results in the hydrogen reionization history consistent with all current observational constraints. Comparing results of the $f_{\rm esc}=0.1$ model and two alternative models shows that the model with a strong luminosity dependence of $f_{\rm esc}$, which assigns high $f_{\rm esc}$ to faint galaxies, results in early reionization inconsistent with observational constraints. However, the model in which $f_{\rm esc}$ follows a universal redshift-independent correlation with the recent maximum specific star formation rate, motivated by the results of the SPHINX galaxy formation simulation, results in the reionization history in good agreement with existing observational constraints, even though this model produces a sizeable ionized hydrogen fraction of $Q_{\rm HII}\approx 0.15-0.2$ at redshifts $z=8-12$. 
Our results show that the relative contribution of faint dwarf galaxies to reionization depends sensitively on assumptions about the escape fraction for galaxies of different luminosities, and that this is the main source of uncertainty in modeling hydrogen reionization.

\end{abstract}


\maketitle


\section{Introduction}
\label{sec:intro}

Ever since \citet{Gunn.Peterson.1965} pointed out that transmission of the rest-frame UV emission in quasar spectra indicates that the intergalactic hydrogen is in a highly ionized state, an ever-increasing body of observations has revealed that the intergalactic gas transitioned from neutral to ionized state around redshift $z\approx 6$ \citep[e.g.,][see \citealt{Robertson.2022,Gnedin.Madau.2022} for reviews]{Becker.etal.2001,Gnedin.2004,Gnedin.Fan.2006}. 
Both observations and cosmological simulations show that this process of {\it reionization} was dominated by the ionizing photons from young star-forming galaxies \citep[e.g.,][]{Gnedin.2000a,Gnedin.Kaurov.2014, Ma.etal.2015,Robertson.etal.2015,Sharma.etal.2016, Madau.2017, Lewis.etal.2023}, with active galactic nuclei (AGNs) playing a key role in helium reionization at $z<4$ and maintaining the intergalactic medium (IGM) ionized at lower redshifts \citep[e.g.,][]{Sokasian.etal.2003}. 

It is still not entirely clear which galaxies dominate the reionization of hydrogen at $z\approx 6$. This is partly because the ionizing photon flux cannot be directly observed, and needs to be estimated with a number of assumptions from the observed UV emission. Besides, the contribution of galaxies with UV absolute magnitudes at $\lambda=1500\,\aa$, or $\Muv >-14$ to the UV flux budget is unconstrained by observations. For example, opposite conclusions have been reached about the relative importance of bright and faint galaxies to the reionization of the Universe, due to different assumptions about ionizing photon contribution from galaxies of different luminosities \citep[e.g.,][]{Finkelstein.etal.2019,Naidu.etal.2020}. 


The relative contribution of different galaxies to the ionizing photon budget depends on the ionizing flux function of galaxies and the dependence of the escape fraction of ionizing radiation, $\fesc$, on galaxy properties. 
Several recent studies have explored the effect of different models for the latter using semi-analytic or semi-numerical models in which $\fesc$ is parameterized as a function of galaxy or halo properties and calibrated to satisfy constraints on the reionization history. In particular, \citet{Seiler.etal.2019} showed that different scalings of $\fesc$, including models tied to halo mass, star formation, and the fraction of gas ejected by stellar feedback, can produce substantially different reionization histories and ionizing-source populations even when calibrated to the same Thomson optical depth \citep[cf. also][]{Hutter.etal.2021,Hutter.etal.2023}. In general, most models find that feedback effects tend to increase $\fesc$ for lower-luminosity galaxies \citep{Hutter.etal.2023, Qin.etal.2021,Mutch.etal.2024}.  

Theoretical models and simulations, likewise, are plagued by uncertainties related to the challenge of correctly modeling the entire spectrum of galaxy masses \citep[e.g.,][]{Gnedin.etal.2008, OShea.etal.2015, Gnedin.2016,  Yue.etal.2016, Kannan.etal.2022}, and face the even more challenging problem of modeling escape of ionizing radiation from galaxies \citep[e.g.,][]{Rosdahl.etal.2022}.

In a recent study, we used a model that reproduces the observed properties of both low-redshift dwarf galaxies and the UV luminosity function of galaxies at $z\approx 5-13$ to investigate the potential contribution of faint galaxies to the UV and ionizing photon budgets \citep{Wu.Kravtsov.2024}. We showed that the contribution of galaxies with $\Muv>-14$ to the UV flux and ionizing photon budget is $\approx 40-60\%$ at $z>7$ and decreases to $\approx 20\%$ at $z=6$ if one assumes that escape fraction of ionizing photons is independent of galaxy luminosities. The contribution of dwarf galaxies is potentially larger if their ionizing photon escape fractions are larger than those of brighter galaxies. These results indicated that dwarf galaxies fainter than the observational limit at $z > 5$ can contribute significantly to the UV flux density and ionizing photon budget.

In this follow-up study, we present calculations of the reionization history of the Universe using the same galaxy formation model and several assumptions about the escape fraction of ionizing photons in galaxies of different luminosity. In the context of previous studies reviewed above, this study contributes calculations with a well-tested galaxy formation model that reproduces both the properties of local dwarf galaxies and the evolution of the UV luminosity function of galaxies at $z>5$. In addition, 
we incorporate in our analysis a model of $\fesc$ based on a recent state-of-the-art SPHINX simulation. 

The calculation of ionizing fluxes of galaxies is done as described in \citet{Wu.Kravtsov.2024}, but we extend our modeling of galaxies to $z\approx 16$, and demonstrate that the model reproduces existing estimates of the UV luminosity function (LF) at $5<z<16$ relevant for robust modeling of reionization. The evolution of the ionized fraction of the intergalactic hydrogen is carried out using the model of \citet[][with some modifications detailed in \citealt{Madau.2017}]{Madau.etal.1999}. We show that assumptions about escape fraction from galaxies are the \textit{largest} uncertainty factor in theoretical modeling of reionization – far larger than modeling the relative contribution of ionizing photons by galaxies of different luminosities, or uncertainties related to the clumping factor of the intergalactic gas. In particular, we show that within a range of reasonable approaches to modeling escape fraction, a wide range of reionization histories can be obtained, including histories in good agreement with existing observational constraints on the hydrogen ionized fraction and Thomson optical depth.

The paper is organized as follows.
We describe the galaxy formation model, and the range of adopted models for the escape fraction of ionizing radiation in Section~\ref{sec:modeling}. We present our main results  in Section~\ref{sec:results}, and discuss their implications in Section \ref{sec:discussion}. Our results and conclusions are summarized in Section~\ref{sec:summary}. We provide parameters for our stochastic star formation model in Appendix~\ref{app:stochastic_params} and fitting functions to both the UV LF and ionizing photon abundance in Appendices~\ref{app:uvlf_fit} and \ref{app:nion_params}.

Throughout this paper, we assume flat $\Lambda$+Cold Dark Matter ($\Lambda$CDM) cosmology with the mean density of matter in units of the critical density of $\Omega_{\rm m}=0.32$, the mean density of baryons of $\Omega_{\rm b}=0.045$, Hubble constant of $H_0=67.11\,\rm km\,s^{-1}\,Mpc^{-1}$, the amplitude of fluctuations within the tophat spheres of $R=8h^{-1}$ Mpc of $\sigma_8=0.82$ (where $h$ is the reduced Hubble constant), and the primordial slope of the power spectrum of $n_{\rm s}=0.95$.

\section{Modeling high-$z$ galaxy formation}
\label{sec:modeling}

The galaxy formation framework we use in this study is applied to predict galaxy population properties for representative samples of model galaxies at all relevant luminosities down to UV absolute magnitudes of $M_{1500}\approx -5$ and redshifts $z\in [5,16]$. This is done using samples of halos that follow the expected halo mass function at each considered redshift and halo mass evolution tracks constructed using an accurate approximation for the halo mass accretion rate, as described in \citet{Kravtsov.Belokurov.2024}. 

The key aspect of the galaxy formation model we use in this study at $z\geq 5$ is that it reproduces observed properties of sub-$\Ls$ galaxies at $z=0$ down to the faintest ultra-faint dwarf galaxies \citep[][]{Kravtsov.Manwadkar.2022,Manwadkar.Kravtsov.2022,Kravtsov.Wu.2023,Pan.Kravtsov.2023}. This agreement is not a guaranty that the model would work at high redshifts. Nevertheless, most galaxies at $z>5$ have dwarf halo virial masses ($\M200c\lesssim 10^{11}\,M_\odot$) similar to that of $z=0$. Therefore, the agreement at $z=0$, as well as the successful forward modeling of $5\leq z \leq 10$ UV LF in \citet{Wu.Kravtsov.2024}, give us more confidence that results may be realistic for higher redshifts. Although the model requires increasing stochasticity of star formation rate with increasing redshift to match the observed UV luminosity function at $z>5$ \citep[][]{Kravtsov.Belokurov.2024}, models with such increased stochasticity are consistent with most observed properties of $z=0$ dwarf galaxies \citep[][]{Pan.Kravtsov.2023}.

\subsection{Halo evolution and star formation model}
\label{subsec:sf_model}

To model the evolution of galaxies over the entire relevant range of galaxy luminosities, we first construct large samples of model halos using the approach described in \citet{Kravtsov.Belokurov.2024}, with specific choices for modeling reionization detailed in \citealt{Wu.Kravtsov.2024}, and adjustments for $z \ge 10$ galaxies in Appendix~\ref{app:stochastic_params}. Briefly, halo samples over a broad range of masses, required to cover the entire relevant range of luminosities, are drawn from a series of cubic volumes of increasing size. Smaller (larger) volumes are used to sample halos of smaller (larger) masses. For each cube halo masses are randomly drawn to follow the halo mass function using the approximation of \citet{Tinker.etal.2008} with the inverse transform sampling method. 
When we construct the overall UV or ionizing radiation luminosity function of galaxies, we generate the samples of halos and the galaxies they host in individual boxes and then stitch the functions in overlapping ranges of halo mass/galaxy UV luminosity where each LF is fully sampled.

To avoid large numerical cost of integrating many evolutionary tracks, we sub-select a random fraction of halos for modeling, given as a function of halo mass $\M200c$ using: 
\begin{equation}
    f(\M200c) = n \left(\frac{\M200c}{10^9\, M_\odot}\right)^\eta,
\end{equation}
where $n\approx 10^{-5}-5\times 10^{-6}$ and $\eta\approx 1.5-2.35$ provide sufficiently large samples of model galaxies to reliably measure their luminosity functions.
When the luminosity function is computed at a given $z$, each model galaxy is weighted by $f^{-1}(\M200c)$. This approach allows us to keep the number of model galaxies reasonably small, while sufficiently sampling galaxies of different luminosities.

Once a halo sample is drawn at a given redshift, $z_{\rm f}$, we use each halo mass as a starting point and construct a halo mass evolution track by integrating the equation of halo mass evolution $\dot{M}_{\rm 200c}=\mu(\M200c)$ back in time to $z_{\rm init}=25$
using an accurate approximation for the {\it average\/} halo mass accretion rate $\dot{M}_{\rm 200c}$ of halos of a given mass $\M200c$, derived using analyses of the mass evolution histories of halos formed in cosmological $\Lambda$CDM simulations:
\begin{equation}
\dot{M}_{\rm 200c} = 0.3606\, M_\odot\, {\rm Gyr}^{-1}\, \M200c(t)^{1.091}\, t^{-1.8},
\label{eq:dmdt}
\end{equation}
where halo mass $\M200c$ is in $M_\odot$ and time $t$ is in Gyrs. The integrated $\M200c(t)$ is then used to model galaxy evolution from $z_{\rm init}$ to a given $z_{\rm f}$. 
The tests of this approximation, especially at $z>5$, and comparisons with other approximations are presented in the Appendix of \citet{Kravtsov.Belokurov.2024}.

Once a given halo mass track is computed, the galaxy evolution track using the model of \citet{Kravtsov.Manwadkar.2022} with modifications to account for stochasticity of star formatin rate \citep{Pan.Kravtsov.2023,Kravtsov.Belokurov.2024}. Specifically, the SFR in the model at time $t_n$ is perturbed as $\dot{M}_{\star,\mathrm{stoch}} = \dot{M}_{\star} \times 10^\Delta $, where $\Delta$ is the a correlated random number drawn from a Gaussian distribution with zero mean and unit variance. 
The number $\Delta$ is modulated to include time correlations by scaling it with $\sqrt{P(k)} = \sqrt{\mathrm{PSD}(f_k)/T}$ \citep[see][for details]{Pan.Kravtsov.2023}, where the power spectral density (PSD) depends on the temporal frequency $f_k$, $T$ is the duration of the galaxy’s evolutionary track, and the corresponding wavenumber is defined as $k = f_k T$. Following \citet{Tacchella.etal.2020} we use the PSD of the form ${\rm PSD}(f) = \sigma^2_\Delta[1+(\tau_{\rm break}f)^\alpha]^{-1}$,
where $\sigma_\Delta$ characterizes the amplitude of the SFR variability over long time scales and  $\tau_{\rm break}$ characterizes the timescale over which the random numbers are effectively uncorrelated. Parameter $\alpha$ controls the slope of the PSD at high frequencies (short time scales). We note that \citet{Pan.Kravtsov.2023} explicitly showed that adding SFR stochasticity improves agreement of the model with observations of local dwarf galaxies.

For each halo track produced as described above, the galaxy formation model is integrated from $z_{\rm init}=25$ to the final redshift $z_{\rm f}=5, 6, 7, 8, 9, 10$, and from $z_{\rm init}=35$ for $z_{\rm f}=11, 12, 13, 14, 16$ to ensure the halo tracks are integrated for enough time, producing the evolution of stellar mass, star formation rate, etc. We tested each of the initial redshift values for convergence of the resulting luminosity and ionizing flux functions. 

\begin{figure*}
   \centering {
   \includegraphics[width=0.98\textwidth]{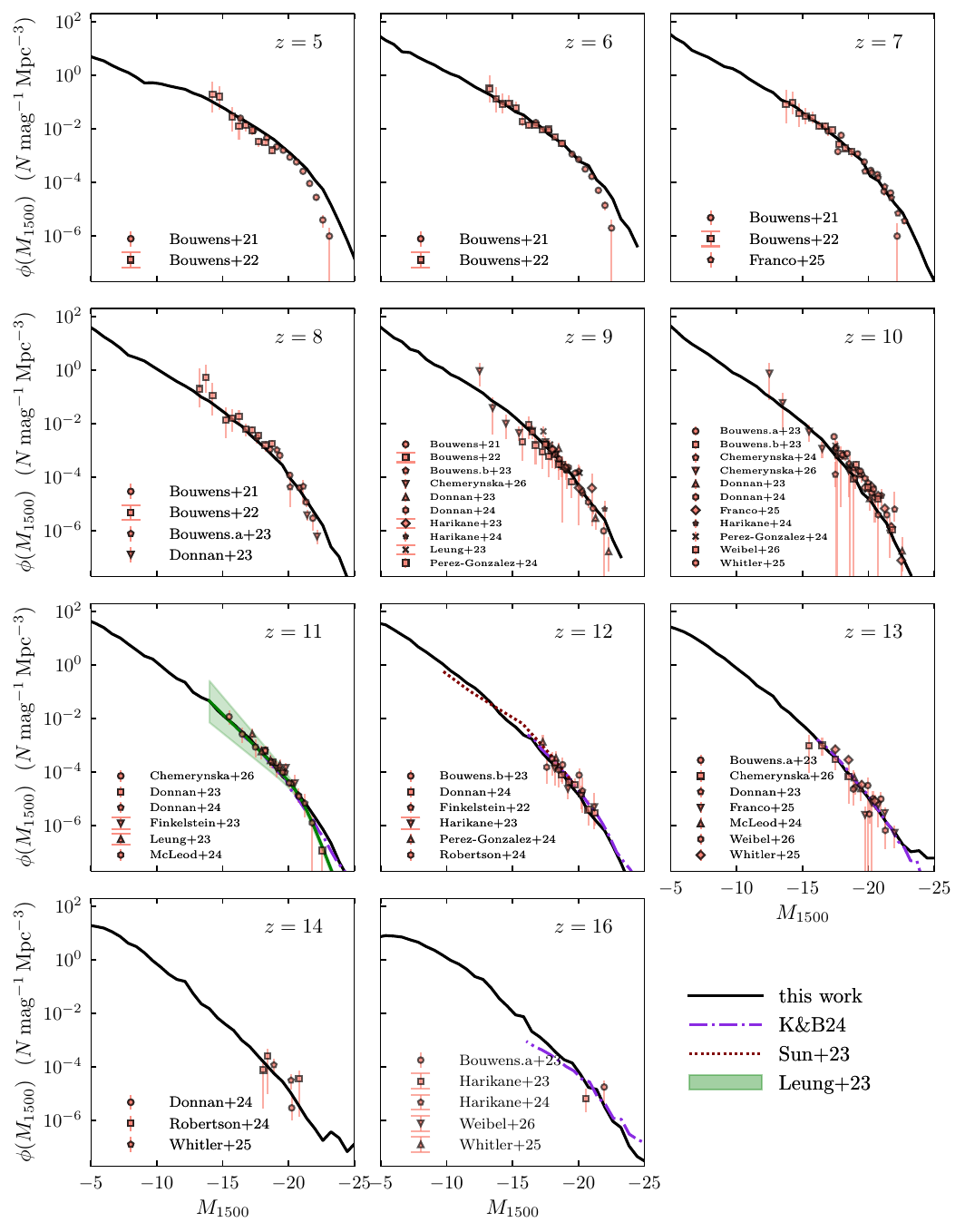}
   }
   \caption{
   Rest-frame UV luminosity function of galaxies in the \GRUMPY model with specific modifications for modeling reionization detailed in \citealt{Wu.Kravtsov.2024, Kravtsov.Belokurov.2024} (see Appendix \ref{app:stochastic_params}), each panel showing a redshift at $z\in[5, 16]$. Effects of dust are not included in the model LFs \citep[see \S 2.3 of][]{Wu.Kravtsov.2024}. The different symbols show observational estimates of the UV LF in recent studies that used HST and JWST observations \citep{Bouwens.etal.2021, Bouwens.etal.2022, Bouwens.etal.2023a, Bouwens.etal.2023b, Chemerynska.etal.2024, Chemerynska.etal.2026, Donnan.etal.2023, Donnan.etal.2024, Finkelstein.etal.2022, Finkelstein.etal.2023, Franco.etal.2025, Harikane.etal.2023, Harikane.etal.2024, Leung.etal.2023, McLeod.etal.2024, Perez.Gonzalez.etal.2024, Robertson.etal.2024, Weibel.etal.2026, Whitler.etal.2025}. Theoretical models are shown in dotted (\citealt{Sun.etal.2023}), dash-dotted (\citealt{Kravtsov.Belokurov.2024}) lines and colored regions \citep{Leung.etal.2023}. Note that before reionization (i.e. $z \gtrsim 6$), the slope of the LF even at the faintest magnitudes remains as steep as the slope at $M_{1500}\approx -14$. For a single-panel comparison of UV LF evolution over redshift, see Fig. 2 of \citet{Wu.Kravtsov.2024}.
   }
   \label{fig:UVLF_obsv}
\end{figure*}

\subsection{Computing UV and ionizing radiation luminosities}
\label{subsec:comp_uv_ion}

The monochromatic luminosity of model galaxies at $\lambda=1500\,{\aa}$ is computed using a tabulated grid of luminosities, $L_{1500}$, for stellar populations of a given age and metallicity using the Flexible Stellar Population Synthesis model v3.0 \citep[FSPS,][]{Conroy.etal.2009,Conroy.Gunn.2010} and its Python bindings, \textsc{Python-FSPS}. The table is then used to construct an accurate bivariate spline approximation to compute $L_{1500}$ for stellar populations of a given age and metallicity. 
We use the table and finely spaced time outputs of the model to compute the integral $L_{1500}$ due to all stars formed by the current time taking into account the evolution of stellar mass and stellar metallicity. 

The emission rate of LyC photons with $\lambda<912\,\aa$,  $\dNion$, is computed using ionizing flux tables from the BPASS version 2.3 models \citep{Byrne.etal.2022}, which take into account the effect of binary stars, and the evolution of stellar mass and metallicity computed by the galaxy formation model. We neglect the effects of dust, which affect only a limited range of bright luminosities at $z\lesssim 7$ \citep[see Fig. 1 in][]{Wu.Kravtsov.2024}.
The UV luminosity and ionizing photon flux functions of model galaxies are then constructed using a weighted contribution of objects from different cubic volumes. We computed the ionizing flux contributions for all galaxies with the monochromatic UV absolute magnitudes at $\lambda=1500\AA$ of $M_{1500}<-5$, for the reasons discussed in \S 2 of our previous paper \citep[][where we refer the reader for other modeling details]{Wu.Kravtsov.2024}:

\begin{equation}
   \langle \dot{n}_{\rm ion}\rangle = \int_{-\infty}^{-5} \,\frac{d\dot{n}_{\rm ion}}{d\Muv}\, d\Muv,
   \label{eq:nion_func}
\end{equation}

The UV luminosity functions produced with this method over a range of absolute magnitudes $-25\lesssim M_{1500}\le -5$ and redshift range $z=5-16$ are shown as black solid lines in Figure~\ref{fig:UVLF_obsv} and compared to existing observational estimates. In our model, the end of reionization \texttt{GRUMPY} parameter is set to $\zrei = 6$ \citep[see §3.2 of][]{Wu.Kravtsov.2024}. As discussed, we adopt a stochastic SFR model in line with \citet{Kravtsov.Belokurov.2024}, with specific values of $\sigma_\Delta$ for each redshift presented in Appendix~\ref{app:stochastic_params}. Best-fit parameters of the modified Schechter function fits to the UV luminosity and ionizing flux functions are also provided in Appendix~\ref{app:uvlf_fit}.

Figure~\ref{fig:UVLF_obsv} shows that the model captures the shape and evolution of the UV luminosity function across the luminosity and redshift range probed by observations. The slow evolution of UVLF amplitude at $z>9$ is mainly due to the assumed increasing {\sl rms} of the stochastic SFR variations with increasing $z$ at these redshifts \citep[see][for details]{Kravtsov.Belokurov.2024}. We quote the values of $\sigma_\Delta$ -- the {\sl rms} of the $\log_{10}\rm SFR$ -- assumed at different $z$, and corresponding {\sl rms} of the UV absolute magnitude, $\sigma_{M_{\rm UV}}$,  in Table~\ref{tab:stoch_params}. Increased stochasticity in higher $z$ galaxies is generally due to their smaller halo masses, which makes them more susceptible to bursts of star formation and violent blowouts of gas, as has been demonstrated in high-resolution simulations of high-$z$ galaxies \citep[e.g.,][]{Sun.etal.2023}. 

Differences between model UVLF and observational estimate at $z\leq 7$ and $M_{1500}\lesssim -20$ are likely due to dust effects that have a similar magnitude at these luminosities and redshifts \citep[see Fig. 1 of][]{Wu.Kravtsov.2024}. In addition, the figure also shows theoretical estimates from \citet{Sun.etal.2023, Leung.etal.2023}, as well as \citet{Kravtsov.Belokurov.2024} which uses the same model as this work for consistency.

Figure~\ref{fig:UVLF_obsv} also shows that the UV luminosity function at luminosities fainter than those probed by observations is quite steep \citep[$\approx-1\sim-2$, depending on the redshift, see Table 1 of ][for more details]{Wu.Kravtsov.2024} illustrating the potentially large contribution of faint galaxies to the UV and ionizing photon budgets. 
Agreement with UV LF estimates at a vast range of redshift $z\in[5, 16]$ at $M_{1500}<-14$, and the fact that the model reproduces properties of $z=0$ dwarf galaxies well \citep{Kravtsov.Manwadkar.2022}, including the luminosity function of Milky Way satellites down to the ultra-faint magnitudes \citep{Manwadkar.Kravtsov.2022}, means we can plausibly expect that the model UV LF can faithfully describe the evolution of the UV luminosities and contribution of ionizing flux from galaxies of luminosities fainter than the observational limit.

\subsection{Models for the escape fraction of ionizing photons}
\label{subsec:fesc_models}

In addition to computing $\dNion$, to compute the reionization history of the Universe we need to make assumptions about the value of the escape fraction of ionizing photons from galaxies ($\fesc$). Such assumptions can be motivated by observations \citep[e.g.,][]{Chisholm.etal.2022,Saldana_Lopez.etal.2023,Topping.etal.2022,Cullen.etal.2023,Lin.etal.2024,Saxena.etal.2024} or theoretical models of ionizing radiation propagation in galaxy formation simulations \citep[e.g.,][]{Wise.etal.2014,Sharma.etal.2016,Kimm.etal.2017,Anderson.etal.2017,Rosdahl.etal.2022}. However, both observational estimates and theoretical modeling of the escape fraction have large associated uncertainties.

Therefore, we adopt several models for $\fesc$ that should reasonably bracket the possible trends. These models range from the simplest constant $\fesc$ assumption to the models motivated by the correlations of $\fesc$ with UV luminosity or specific star formation rate from observations and simulations. 

\textbf{1.} In the first model, we assume a \textit{constant} $\fesc=0.1$ at all redshifts and luminosities. In \citet{Wu.Kravtsov.2024} we adopted $\fesc=0.05$, chosen so that the escaping ionizing photon rate density is comparable to the estimate required to keep the IGM ionized \citep{Madau.etal.1999}. Here we instead fully integrate the ionzing history (\S~\ref{subsec:hydrogen_ion_modeling}), and find $\fesc=0.1$ agrees better with observational constraints, comparable to those adopted or inferred by other models \citep[e.g.,][]{Hutter.etal.2021, Mutch.etal.2024, Ferrara.etal.2025}.

\textbf{2.} In the second model, we adopt a strong dependence of $\fesc$ on galaxy luminosity for galaxies with $-21<\Muv<-15$: $\fesc^{\rm eff} = \fesc(\Muv = -21) \times 10 ^ {0.62(\Muv + 21)}$, where $\fesc(\Muv = -21) = 1.91\times 10^{-4}$ normalized the function so that $\fesc$ would increase to 1 towards the faintest galaxies. At $\Muv>-15$ is kept constant at $\fesc=1$.
This model approximates the evolution of $\fesc$ and in simulations \citet{Anderson.etal.2017} and evolution of 
$\xiion$ deduced for observed galaxies by \citet{Simmonds.etal.2024a}.  This toy model illustrates how different the results would be in the case of a strong increase of $\fesc$ with fainter luminosities. Note, however, that the luminosity dependence in this model is likely too strong and overestimates the contribution of dwarf galaxies to the ionizing budget, as it assumes that both $\fesc$ and $\xiion$ increase with decreasing luminosity, while observations indicate that $\fesc$ and $\xiion$ anti-correlate in high-$z$ galaxies \citep{Saxena.etal.2024}, such that their product does not depend strongly on luminosity.

\textbf{3.} Our third model for $\fesc$ is based on its correlation with specific star formation rate (sSFR) of galaxies found in the \texttt{SPHINX20} simulations of galaxy formation. Specifically, \cite[][see their Fig. 12 ]{Rosdahl.etal.2022} found that in these galaxies $\fesc$ increases with increasing {\it maximum} $\ssfr10$ estimated over 50 Myr before the current epoch:
\begin{equation}
    {\rm sSFR_{\rm max,10}} = \max_{50\,\rm Myr}{\ssfr10},
\end{equation}
where
\begin{equation}
    {\ssfr10}= \frac{{\rm SFR}_{10}}{M_\star} = \frac{M_{\star}({\rm age < 10\, Myr})/{10\,\rm Myr}}{M_\star},
    \label{eq:sSFR}
\end{equation}
and $M_\star$ is the current stellar mass of the galaxy, and SFR$_{10}$ is the star formation rate averaged over the past 10 Myr. Observationally, this timescale roughly corresponds to SFRs determined by H$\alpha$. Importantly, \citet{Rosdahl.etal.2022} showed that dependence of $\fesc$ on ${\rm sSFR_{\rm max,10}}$ is similar at different redshifts (see their Fig. 15, also reproduced in Figure~\ref{fig:SPHINX_sSFR} below). Thus, to model such dependence, we only need a single parametrization of the $\fesc(\ssfr10)$.

Physically, this correlation likely reflects a link between escape fraction and starburst-induced feedback, as pointed out and discussed in many previous works \citep{Heckman.etal.2011, Sharma.etal.2017, Faucher-Giguere.2020}, although the processes shaping $\fesc$ remain a matter of ongoing debate. Our third model thus is an example of a simulation-motivated scenario implementing such correlation of $\fesc$ with ${\rm sSFR_{\rm max,10}}$.

Analyzing the distributions of $\fesc$ in the SPHINX simulations as a function of ${\rm sSFR_{\rm max,10}}$, we found that it can be well approximated by a random variable $x=\log_{10}\fesc$ drawn from a two-component (bi-modal) mixture distribution function, where the first model is described by the skew-normal distribution 
\begin{equation}
p_1(x\vert\xi,\omega) = \frac{2}{\omega\sqrt{2\pi}}\,e^{-\frac{(x-\xi)^2}{2\omega^2}}\int_{-\infty}^{\alpha\frac{x-\xi}{\omega}}\frac{1}{\sqrt{2\pi}}e^{-\frac{t^2}{2}}dt     
\end{equation}
and the second model by the normal distribution
\begin{equation}
p_2(x\vert\mu,\sigma)=\frac{1}{\sigma\sqrt{2\pi}}e^{-\frac{(x-\mu)^2}{2\sigma^2}}.
\end{equation}
In the approximation we fix $\alpha=-3$ and $\omega=1.6$ and adopt the following dependencies for $\xi$, $\mu$, $\sigma$ on $\log_{10}{\rm sSFR_{\rm max,10}}$:
\begin{eqnarray}
\xi &=& 1.7(\log_{10}{\rm sSFR_{\rm max,10}}-2) - 0.7\nonumber\\
\mu &=& \log_{10}{\rm sSFR_{\rm max,10}}-2.3\\
\sigma &=& 0.45 - 0.35(\log_{10}{\rm sSFR_{\rm max,10}}-2).\nonumber
\end{eqnarray}
We assume that the distribution of galaxies is produced by the equal fractions (50\%) of galaxies drawn from the two components). All drawn $\fesc$ values are capped at $\fesc=1$. 

Figure~\ref{fig:SPHINX_sSFR} compares the $\fesc$ distributions of the SPHINX galaxies and values produced using this approximation. The solid lines with different colors show the medians of $\fesc$ distributions of \texttt{SPHINX} galaxies at different redshifts, while the shaded regions show the $68$th and 84th percentiles of the distribution of $\fesc$ (at $z=6$, the distributions at other $z$ are similar). The thick dashed line shows the median, and the thinner dotted lines show the $68$th and 84th percentiles of the distribution for the $\fesc$ values produced using this approximation for the same ${\rm sSFR_{\rm max,10}}$ values. The figure shows that the approximation matches the SPHINX simulation results at  $\log_{10}{\rm sSFR_{\rm max,10}}>0$ reasonably well.
Our approximation underestimates $\fesc$ at ${\rm sSFR_{max,10}}\lesssim1\,{\rm Gyr^{-1}}$ where the median $\fesc$ of SPHINX galaxies increases: a non-monotonic trend that our linear parametrization does not capture. However, such galaxies formed little stellar mass over the preceding 50 Myr and therefore should produce few ionizing photons, while their $\fesc$ in SPHINX is still $\lesssim10^{-2}$. Their contribution to the escaping ionizing photon budget is negligible, and a more complex parametrization in this regime does not affect our results.

In the third $\fesc$ model we use this approximation of SPHINX galaxies to assign $\fesc$ values to model galaxies in our calculation using the ${\rm sSFR_{\rm max,10}}$ values computed from the star formation history of each model galaxy, accounting for SFR stochasticity. 
Our calculations showed that when applying this model to calculate hydrogen ionization history, we found reionization occurring too late ($z\lesssim 5$). This is primarily because escape fraction in SPHINX simulations have not converged with resolution, so a higher resolution would yield higher values of $\fesc$, even though star formation rates \textbf{have} converged and the qualitative $\fesc$ trends are consistent \citep[see Fig. 19 in][]{Rosdahl.etal.2022}. Indeed, the SPHINX20 simulation itself also produces a somewhat late reionization.
To address this, we modify our model such that $\fesc$ of each galaxy retains the same dependencies on ${\rm sSFR_{\rm max,10}}$, but is boosted by a factor of two. We note here that boosting a higher $\fesc$ or even star formation efficiency is a somewhat common approach adopted by comparable cosmological simulations \citep[e.g.][]{Acharya.etal.2025} to combat the underestimation of ionizing photons due to insufficient resolution. We show later that this adjustment provides a good match to existing observational constraints on reionization, and therefore adopt this version as our fiducial third $\fesc$ model. We also note that a boost of $\fesc$ by such a factor implies a significant change in the distribution of neutral hydrogen in the simulations. At the same time, one could think about this change as a factor of two change of the uncertain intrinsic ionizing flux of stellar populations due, for example, to a different model of binary star population. $\fesc$ values are always capped at 1.

We propagate the ionizing photons from the aforementioned three $\fesc$ models with the method outlined in Section~\ref{subsec:comp_uv_ion}, and present ionizing histories in Section~\ref{subsec:hydrogen_ion_modeling}.

\begin{figure}
   \centering {
   \includegraphics[width=0.49\textwidth]{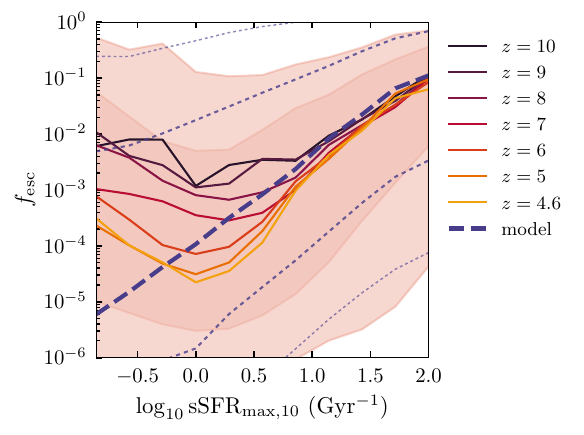}
   }
   \caption{
   The $\fesc$ distribution for galaxies in the \texttt{SPHINX} simulation, binned by maximum of the logarithm of the 10 Myr-averaged specific star formation rate $\ssfr10$ measured over the past 50 Myr. The solid curves represent median $\fesc$ values for different $\log_{10}{\rm sSFR_{\rm max,10}}$ at different redshifts, and shaded regions correspond to 68th and 84th percentiles for SPHINX galaxies at $z=6$ (the distributions at other $z$ are similar). The dashed thick line represents the median, and the thinner dotted lines represent the 68th and 84th percentiles of the distribution for the same galaxies, but with $\log_{10}(\fesc)$ modeled using the approximation described in \S~\ref{subsec:fesc_models}.
   }
   \label{fig:SPHINX_sSFR}
\end{figure}


\subsection{Modeling hydrogen ionization history}
\label{subsec:hydrogen_ion_modeling}

To quantitatively model the hydrogen ionization history, we construct an ionizing flux $\dot{n}_{\rm ion}(\Muv)$ as a function of $\Muv$, similarly to how we estimate model UV LF in \citet{Wu.Kravtsov.2024}. We then use a polynomial fit to model the redshift evolution of the best-fit Schechter functional form parameters over $5\leq z\leq 16$, also described in the previous work. The method is outlined here again in Appendix~\ref{app:uvlf_fit}, with the resulting parameters presented in Appendix~\ref{app:nion_params}.

With this parameterization of the time evolution of $\dot{n}_{\rm ion}(\Muv)$ we can compute $\dot{n}_{\rm ion}(<M_{1500}^{\rm lim}, z)$, where we integrate ionizing photons at each redshift down to the luminosity limit $\Muv < -5$ on the faint end. This sums $\dNion$ flux contribution from all dwarf galaxies in our model, including those not currently observable by JWST at $\Muv > -13$. We also integrate to three different luminosity limits in the discussion Section~\ref{subsec:gal_contribution}, to investigate ionizing photon contributions from galaxies of different luminosities.

We use this function to solve for the volume-averaged hydrogen ionized fraction $Q(t)$  \citep{Madau.etal.1999}, rewriting it in a redshift-dependent form $Q(z)$ using $dt = - dz[H(z)(1+z)]^{-1}$:

\begin{align}
    \frac{dQ}{dz} &= - \frac{1}{H(z)(1+z)}\frac{dQ}{dt}\nonumber\\ &= - \frac{1}{H(z)(1+z)}\left(\frac{\langle \dnion \rangle}{\langle n_{H} \rangle} - \frac{Q}{\bar{t}_{\rm rec}} \right),
    \label{eq:dqdz}
\end{align}
where $\langle \dnion \rangle=\dot{n}_{\rm ion}(<\Muv, z)$ is the ionizing photon production rate per unit proper volume, and $\langle n_{H} \rangle = 1.89\times 10^{-7} (1+z)^3 {\rm cm}^{-3}$ is the cosmological mean proper hydrogen density. We compute the ``effective'' recombination timescale $\bar{t}_{\rm rec}$ for $\rm H_{II}$ regions in the IGM using equations and assumptions of \citet{Madau.etal.2024} as:

\begin{equation}
    1/ {\bar{t}_{\rm rec}} \equiv (1 + \chi) \alpha_{B}(T_0) \langle n_{H} \rangle C_R,
    \label{eq:trec}
\end{equation}

where $\alpha_B=2.577\times 10^{-13}\rm cm^3s^{-1}$ is the combination coefficient computed with the fixed temperature of ionized gas of $T_0 = 10^4 \rm K$, $\chi \equiv Y/4X$, which assumes that helium is singly ionized at the same time as hydrogen, and doubly ionized only at later times.

We consider three redshift-dependent models of the IGM clumping factor $C_R(z)$.  Following \citet{Madau.etal.2024}, our fiducial model is a parameterization $C_R = 9.25 - 7.21\log_{10}(1+z)$ calibrated from  the radiation-hydrodynamic simulations of ionization inhomogeneities of \citet{Finlator.etal.2009} and \citet{Chen.etal.2020}. We also consider two other models that bracket the upper and lower range of $C_R(z)$ in different simulations \citep[e.g.][]{Pawlik.etal.2009, Finlator.etal.2012, Shull.etal.2012,Kaurov.Gnedin.2015} shown in Fig. 2 of \citet{Davies.etal.2024b}. Specifically, the model that brackets the simulation results from above has $C_R$ changing from 3 at $z>10$ to 10 at $z \lesssim 6$: $C_R=\min\left[10,3 + z^{-2.5}\exp(-0.3z + 7.4)\right]$, while the model that brackets simulation results from below is $C_R=8 - 6.5 \log_{10}(1+z)$. 

These three models of $C_R(z)$ are plotted in Figure~\ref{fig:clump_z} of the Appendix~\ref{app:model_plots}. The effect of different clumping factor model choices on the hydrogen reionization history is discussed in Section~\ref{subsec:clumping_factor}, while all other calculations use the fiducial model.

\section{Results}
\label{sec:results}

\begin{figure}
    \centering{
    \includegraphics[width=0.49\textwidth]{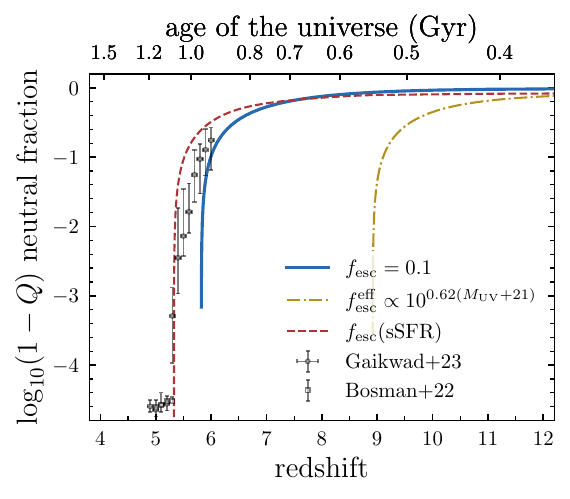}
    }
    \caption{
        Reionization histories represented by the average neutral fraction $1 - Q$ of the intergalactic medium as a function of redshift and cosmic time. The curves show predictions from integrating Eq.~\ref{eq:dqdz} with three models for the ionizing photon escape fraction: a constant global $\fesc = 0.1$ (blue); a model in which $\fesc = \fesc(\Muv = -21) \times 10 ^ {0.62(\Muv + 21)}$ increases monotonically for fainter galaxies (yellow), where $\fesc(\Muv = -21) = 1.91\times 10^{-4}$; and lastly the model where $\fesc$ depends on specific star formation rate (red, see \S~\ref{subsec:fesc_models}). All models integrate galaxies at each redshift across the full luminosity range.
        Points show Ly$\alpha$ forest estimates of the volume-averaged neutral fraction from \citet[][circles]{Gaikwad.etal.2023} and \citet[][squares; no self-shielding]{Bosman.etal.2022}.
    }
    \label{fig:log1Qz_fesc}
\end{figure}

\subsection{Model hydrogen ionization history compared to observations}
\label{subsec:results_ionization_hist}

We integrate Equation~\ref{eq:dqdz} to a given $z$ starting from a fully neutral universe ($Q=0$) at $z_i = 16$. Figure~\ref{fig:log1Qz_fesc} shows the resulting reionization histories represented by the average neutral fraction $1 - Q$ of the intergalactic medium with redshift and cosmic time, with the neutral fraction plotted on a logarithmic scale. This is done to emphasize the reionization epoch, $\zrei$ at which the bulk of the intergalactic medium is ionized. The figure compares the three models for the escape fraction adopted in our study (see Section~\ref{subsec:fesc_models}): a constant global $\fesc = 0.1$, the strongly luminosity-dependent $\fesc^{\rm eff}$ which increases linearly towards fainter galaxies, and the sSFR-dependent $\fesc$ model motivated by the SPHINX galaxy formation simulation. All models include the ionizing flux contribution of all galaxies down $\Muv<-5$.

Figure~\ref{fig:log1Qz_fesc} illustrates the high sensitivity of the reionization history to assumptions about the ionizing photon escape fraction. The model with luminosity-dependent escape fraction (yellow) reionizes around $z \sim 9$. This is too early to be compatible with observations (see below) and is consistent with conclusions of \citet{Munoz.etal.2024} that the Universe would be reionized too early if escape fractions increases rapidly with decreasing galaxy luminosity. \footnote{We note that the tension found by \citet{Munoz.etal.2024} relied on the $\xi_{\rm ion}$ estimates of \citet{Simmonds.etal.2024a} and is reduced with the updated estimates of \citet{Simmonds.etal.2024c}.}

In the $\fesc = 0.1$ model (blue) the universe reionizes at $z\sim 6$. The sSFR-dependent escape fraction model produces a reionization history quite similar to the $\fesc = 0.1$ model. This shows that very different models of escape fraction may produce \textit{similar} reionization histories.

Figure~\ref{fig:log1Qz_fesc} also shows the volume-averaged neutral hydrogen fraction at $4.9\leq z\leq6$ inferred from Ly$\alpha$ forest observations in \citet{Bosman.etal.2022, Gaikwad.etal.2023}. The $\fesc({\rm sSFR})$ model is broadly consistent with these estimates, while in the $\fesc=0.1$ model the neutral fraction falls below the inferred values at $z\approx6$, with reionization completing somewhat earlier than these measurements indicate. We note that a quantitative comparison with Ly$\alpha$ forest constraints requires modeling spatial inhomogeneity of the ionizing background and scatter along different lines of sight, which is beyond the scope of our model.

\begin{figure}
   \centering {
   \includegraphics[width=0.49\textwidth]{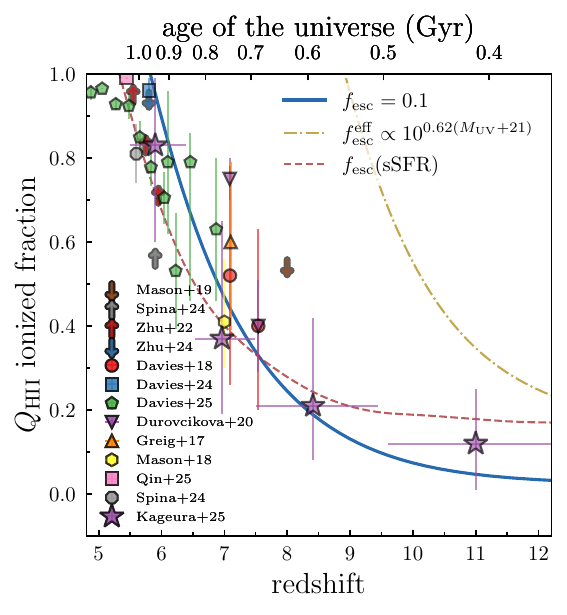}
   }
   \caption{
        Hydrogen ionized fraction $Q_{\rm HII}(z)$ with redshift and cosmic time. The same curves representing $\fesc$ models are shown as in Fig.~\ref{fig:log1Qz_fesc} with the same colors. All models integrate galaxies at each redshift across the full luminosity range down to $\Muv < -5$. Observational measurements (points) and limits (arrows) are plotted from a compilation of Ly$\alpha$, QSO, and IGM temperature measurements \citep[e.g.,][]{Davies.etal.2018, Davies.etal.2024, Davies.etal.2025, Durovcikova.etal.2020, Durovcikova.etal.2024, Greig.etal.2017, Kageura.etal.2025, Mason.etal.2018, Mason.etal.2019, Qin.etal.2025, Spina.etal.2024, Zhu.etal.2022, Zhu.etal.2024}.
    }
   \label{fig:Qz_fesc}
\end{figure}

Figure~\ref{fig:Qz_fesc} compares reionization histories in these three models to observational constraints from observed Ly$\alpha$, QSO, and IGM temperature measurements \citep[e.g.,][]{Davies.etal.2018, Davies.etal.2024, Davies.etal.2025, Durovcikova.etal.2020, Durovcikova.etal.2024, Greig.etal.2017, Kageura.etal.2025, Mason.etal.2018, Mason.etal.2019, Spina.etal.2024, Zhu.etal.2022, Zhu.etal.2024}.

The figure shows that the model with $\fesc = 0.1$ and the simulation-based sSFR-dependent $\fesc$ model results are in good agreement with measurements of the IGM neutral fraction, while the model in which $\fesc$ increases for fainter luminosities greatly overestimates the ionized fraction estimates at $z<10$. Although not shown in the figure, we note that the shape of the $Q(z)$ line for the $\fesc = 0.1$ model is also in good agreement with the parameterization: $Q = [(z_{\rm early} - z) / (z_{\rm early} - z_{\rm end})]^\alpha$ where $\alpha = 4$; $z_{\rm early} = 16$ is the redshift around which the first emitting sources form, and $z_{\rm end}$ is taken at the redshift where reionization draws to a close ($Q = 0.99$, $z = 5.77$), found to be a good fit to the CMB results in \citet{Planck.coll.2016}.

Interestingly, $Q(z)$ in the simulation-based sSFR-dependent $\fesc$ model is similar to $\fesc = 0.1$. Despite this, the reionization history shows a sizeable fraction of ionized hydrogen – $Q_{\rm HII}\approx 0.15\sim0.2$ – is produced at higher redsfhits ($z>10$). Current observational constraints do not exclude such a scenario; and as we show below, such reionization history is consistent with the integrated optical depth constraints from the cosmic microwave background (CMB) measurements.

\subsection{Thomson optical depth}
\label{subsec:thomson_optical_depth}

Another key constraint on reionization histories is the Thomson electron optical depth due to the scattering of the CMB photons by free electrons produced during reionization, which can be computed as \citep{Kuhlen.Faucher-Gigu`ere.2012, Robertson.etal.2015, Robertson.2022}:
\begin{equation}
    \tau(z) = c \sigma_T \langle n_H \rangle \int_0^z dz' \frac{(1+z')^2}{H(z')} \left[1 + \frac{\eta Y}{4X} \right] Q_{\mathrm{HII}}(z'),
    \label{eq:tauz}
\end{equation}
where $c$ is the speed of light, $\sigma_T$ is the Thomson scattering cross section, $\langle n_H \rangle$ is the mean proper hydrogen density, and $Q_{\mathrm{HII}}(z')$ is the volume-averaged ionized fraction computed in Section~\ref{subsec:results_ionization_hist}. The pre-factor accounts for free electrons contributed by helium at different ionization states. We adopt the assumption that helium is fully ionized ($\eta = 2$) at lower redshifts $z < 4$, and singly ionized ($\eta = 1$) at higher redshifts.

\begin{figure}
   \centering {
   \includegraphics[width=0.49\textwidth]{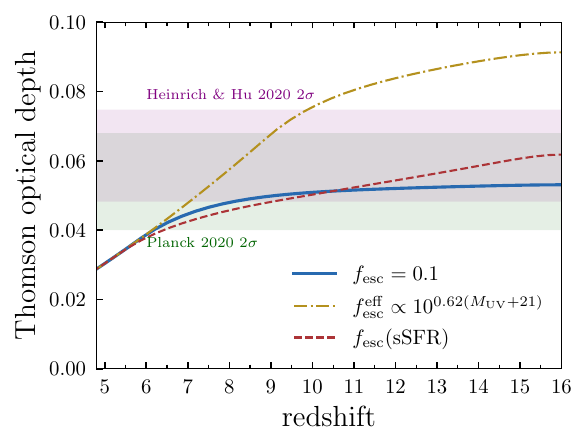}
   }
   \caption{
        Thomson optical depth $\tau(z)$ as a function of redshift for the three ionizing escape fraction models: a constant $\fesc = 0.1$ integrated down to $\Muv < -5$ (blue), a model in which a model in which $\fesc$ increases monotonically for fainter galaxies, and is normalized to 1 for the faintest galaxies (yellow), and the $\fesc$ model depending on specific star formation rate (red). The colored bands shows the $2\sigma$ range of the opacity constraints measured by \citet{Planck.coll.2020} and \citet{Heinrich.Hu.2021}'s reanalysis.
    }
   \label{fig:tauz_fesc}
\end{figure}

Figure~\ref{fig:tauz_fesc} shows the $\tau(z)$ for the same three $\fesc$ models considered above (see Section~\ref{subsec:fesc_models}), which account for the contribution of model galaxies down to $\Muv < -5$. The predicted optical depths are compared to the $\tau = 0.054 \pm 0.014\ (2\sigma)$ constraints from the \citet{Planck.coll.2020} measurements, as well as $\tau = 0.0626^{+0.0122}_{-0.0144} (2\sigma)$ from \citet{Heinrich.Hu.2021}'s more recent PC analysis of the same Planck data.

The figure shows that both the $\fesc = 0.1$ and sSFR-based $\fesc$ models are  consistent with the CMB constraints. The optical depth increase with redshift is similar in both model, but the integral $\tau$ is somewhat higher in the latter model due to the ionized hydrogen fraction tail at $z>10$ in this model, seen in Figure~\ref{fig:Qz_fesc}. In contrast, the model with luminosity-dependent $\fesc$ requires Thomson optical depth that is too large, thanks to the early reionization due to the large ionizing flux contribution of faint galaxies with high escape fractions. Together with the IGM neutral fraction constraints discussed in Section~\ref{subsec:results_ionization_hist}, it is clear that a monotonically increasing $\fesc$ overestimates the dwarf contribution to the reionization.

\section{Discussion}
\label{sec:discussion}

\subsection{Contribution from galaxies of different luminosities}
\label{subsec:gal_contribution}

Comparison of reionization histories and implied integral Thomson optical depth for CMB photon scattering in different models presented above highlights the high sensitivity of the model predictions to assumptions about the escape fraction dependence on galaxy properties.

The contribution of galaxies of different luminosities towards the ionizing photon budget is still somewhat uncertain \citep[see, e.g.,][]{Finkelstein.etal.2019,Naidu.etal.2020,Wu.Kravtsov.2024}. However, we show in this section that the effect of including galaxies of luminosity fainter than the observation limit is considerably smaller than the effect of different assumptions about $\fesc$ – at least for the model employed in our study, which reproduces properties of dwarf galaxies and their UV luminosities at $z\approx 7$, as indicated by their star formation histories \citep[see][]{BoylanKolchin.etal.2015,Weisz.BoylanKolchin.2017,Wu.Kravtsov.2024}

\begin{figure}
   \centering {
   \includegraphics[width=0.49\textwidth]{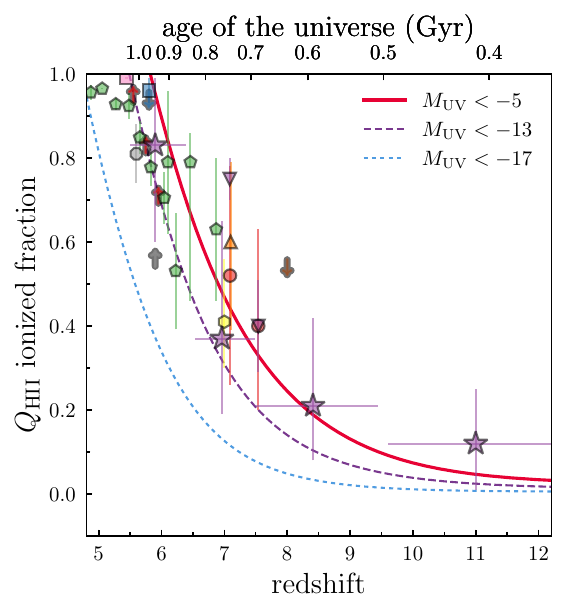}
   }
   \caption{
        Hydrogen ionized fraction $Q_{\rm HII}(z)$ with redshift and cosmic time, assuming a fixed global escape fraction $\fesc = 0.1$ and integrating over different faint-end limits of the galaxy UV magnitude: $\Muv < -17, -13,\ {\rm and} -5$. Also shown are the same set of observational constraints as in Fig.~\ref{fig:Qz_fesc}, from a compilation of Ly$\alpha$, QSO, and IGM temperature measurements.
    }
   \label{fig:Qz_muv}
\end{figure}

For example, Figure~\ref{fig:Qz_muv} shows reionization histories in the $\fesc = 0.1$ model accounting for only the ionizing photons produced by bright galaxies ($\Muv<-17$), by galaxies with $\Muv<-13$, and by all model galaxies $\Muv<-5$. We chose the constant $\fesc=0.1$ model, given that it is in good agreement with all of the observational constraints in the case when we account for all galaxies ($\Muv<-5$, see Figure~\ref{fig:Qz_fesc}). 
The figure shows that including fainter galaxies accelerates the reionization process, in agreement with previous calculations (see, e.g., $\nion$ comparisons in Fig. 6 of \citet{Wu.Kravtsov.2024}). This is especially prominent at lower redshifts $z \lesssim 8$ where including dwarf galaxies with $\Muv > -13$ boosts the ionized fraction by almost 0.1, whereas the difference is not as significant at higher redshifts. 
Nevertheless, if the contribution of dwarf galaxies to the ionizing budget were neglected, to obtain the reionization history required to match observational constraints would only require increasing escape fraction from $\fesc=0.1$ to $\fesc\approx 0.15$ \citep[see, e.g.,][]{Robertson.2022}. 

The uncertainties in the overall contribution of faint galaxies to the ionizing photon production thus have a much smaller effect than the uncertainty in the relative escape fractions in galaxies of different luminosity (and other properties). This uncertainty is generic and is unlikely to be eliminated. Although attempts to characterize  $\fesc$ scaling with galaxy properties in high-resolution simulations continue \citep[e.g.,][]{Wise.etal.2014,Kimm.etal.2017,Anderson.etal.2017,Rosdahl.etal.2022}, the fidelity of each specific simulation is uncertain. 

This reflects the daunting nature of the task: $\fesc$ depends on the details of formation and spatial distribution of massive stars relative to the spatial distribution of neutral gas on an enormous range of scales \citep[see, e.g.,][]{Howard.etal.2017,Howard.etal.2018,Menon.etal.2025} and can thus be easily affected both by the insufficient fidelity in modeling star formation and feedback processes and by resolution effects. 

At the same time, observational estimates of $\fesc$ have their own uncertainties and a set of challenges. First, detailed measurements of escape fraction are difficult and are generally only available for lower redshift galaxies, not for the galaxies responsible for reionizing the Universe. Second, escape fractions for faint galaxies are inherently difficult or impossible to measure. Third, although some galaxy properties were found to correlate with $\fesc$ and thus were proposed as proxies for its estimates \citep[e.g.,][]{Chisholm.etal.2022}, simulation-based tests indicate that 
such observational proxies may be biased or have a generally large scatter \citep[e.g.,][]{Choustikov.etal.2024, Choustikov.etal.2024b}. 

Nevertheless, the utmost sensitivity of the reionization history modeling on the modeling of the ionizing photon escape fractions dictates that a combination of simulation-based inference and observations should be used to make progress. For example, despite the inherent challenges and uncertainties of the simulations discussed above, simulations can be used to test observational diagnostics or suggest the most promising properties of galaxies as proxies and related parametrizations of $\fesc$. 
Promising efforts of this kind have recently been undertaken \citep[e.g.,][]{Choustikov.etal.2025}

\subsection{Reionization history dependence on the IGM clumping factor}
\label{subsec:clumping_factor}

The hydrogen reionization history is additionally sensitive to the assumed clumping factor, $C_R$, which affects the global hydrogen recombination rate. To illustrate the sensitivity of the reionization history to $C_R$, we show such histories for our fiducial model for $C_R(z)$ and the two models bracketing $C_R(z)$ in different simulations (see Section~\ref{subsec:hydrogen_ion_modeling} for details).

\begin{figure}
   \centering {
   \includegraphics[width=0.49\textwidth]{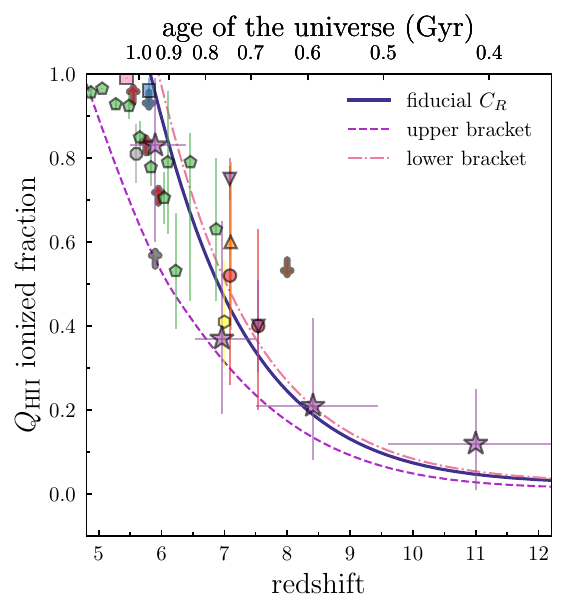}
   }
   \caption{
        Hydrogen ionized fraction $Q_{\rm HII}(z)$ as a function of redshift and cosmic time, assuming a fixed global escape fraction $\fesc = 0.1$ and accounting for the ionizing photons from all galaxies down to $\Muv < -5$. Reionization history is computed with a fiducial model for the IGM clumping factor model as a function of redshift $C_R(z)$ (blue) and two models bracketing results for $C_R(z)$ from a number of different simulations from above (purple) and below (pink line). The models are described in \S~\ref{subsec:hydrogen_ion_modeling} and \ref{subsec:clumping_factor}. Also shown are the same set of observational constraints as in Fig.~\ref{fig:Qz_fesc}. The fiducial model and lower bracket models are close to each other and thus have similar reionization histories, while in the upper bracket model the Universe reionizes somewhat later.
    }
   \label{fig:Qz_clump}
\end{figure}

Figure~\ref{fig:Qz_clump} shows the reionization histories for these three $C_R$ models. As expected, a higher clumping factor delays reionization, since it requires a larger number of ionizing photons to fully reionize the Universe. Overall, however, the effect of different $C_R$ prescriptions on the reionization history is modest and is much smaller than effect of choosing different models for $\fesc$, or even the effects of including galaxies of different luminosities discussed above. This is true even though the range of values allowed between the models is likely larger than the actual range of plausible $C_R$ values, as indicated by the study by \citet{Gnedin.2024}, in which the clumping factor was modeled based on the combination of cosmological simulations and results of very high-resolution small-box simulations.

\subsection{Comparisons with previous studies}
\label{subsec:comparisons}

Escape fraction, ionizing model flux, and reionization of the Universe have been modeled in a number of recent studies. For example, \citet{Ferrara.etal.2025} present an analytical model of galaxy population and reionization is presented. The ionizing photon escape fraction is modeled using an empirically-motivated model and depends on the instantaneous specific star formation rate. Although this is different from our fiducial model, in which $\fesc$ depends on the {\it maximum} sSFR over the past 50 Myr, where SFR is measured over the period of 10 Myr, we find a similar trend of increasing $\fesc$ with increasing redshift as in Figure 1 of \citet{Ferrara.etal.2025}. Although the median $\fesc$ in our model is two orders of magnitude smaller than in that study, the $\fesc$ in our model has a large scatter and the $+1\sigma$ and $+2\sigma$ regions above the median contain galaxies with comparable $\fesc$ to the values in the \citet{Ferrara.etal.2025} model. We note also that their model predicts $\fesc\approx 0.03-0.1$ at $z\approx 6-8$ during the last stages of reionization, which is consistent with the typical values of $\fesc$ required to reionize the Universe at these redshifts in our model \citep[][]{Wu.Kravtsov.2024}. Thus, even though the models use different definitions of sSFR and different modeling of $\fesc$, the actual values of escape fraction and resulting reionization histories are qualitatively similar. 

\begin{figure}
   \centering {
   \includegraphics[width=0.49\textwidth]{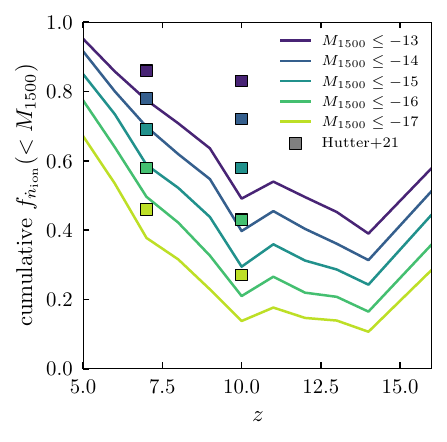}
   }
   \caption{
        Cumulative fraction of the total ionizing photon production rate $\dot{n}_{\rm ion}$, contributed by \GRUMPY galaxies brighter than given UV magnitude limits as a function of redshift. Different curves show limits from $M_{1500}\leq-13$ to $\leq-17$, normalized against the ionizing photon of all model galaxies integrated down to $M_{1500}\leq-5$. Square markers show corresponding fractions in \citet[][``photoionization'' model, Table 1]{Hutter.etal.2021} at $z=7,10$ with colors matching magnitude limits. The comparison shows a stronger contribution to the ionizing photon budget from fainter galaxies in our model.
    }
   \label{fig:fracnion_Muv}
\end{figure}

A decomposition of the ionizing photon budget by galaxy luminosity and host halo mass was carried out by \citet{Hutter.etal.2021}, who explored a suite of seminumerical models. Three of these (`photoionization', `strong heating', and `Jeans mass') assume a constant $\fesc$ and differ only in the strength and timing of radiative feedback, while `early heating' links $\fesc$ to the fraction of gas ejected from the galaxy, such that $\fesc$ increases towards lower halo masses.
Figure~\ref{fig:fracnion_Muv} compares the cumulative fraction of $\dot{n}_{\rm ion}$ in our fiducial $\fesc({\rm sSFR})$ model, contributed by galaxies brighter than a given UV magnitude, with the fractions tabulated in their Table 1 at $z=7$ and $z=10$. 

Qualitatively, the trends in our model are similar to those in \citet{Hutter.etal.2021}. However, the fractions of ionizing photons contributed by galaxies brighter than a given magnitude are systematically lower in our model due to a considerable contribution of faint dwarf galaxies to the ionizing photon budget.  We note that the \citet{Hutter.etal.2021} calculations include resolved halos down to $M_{\rm h} \simeq 10^{8.2}\,M_\odot$ whereas we normalize to all model galaxies down to $M_{1500}=-5$, which are hosted in lower mass halos.  We also note their model values span a wide range at fixed magnitude limit: at $z=7$ the fraction contributed by galaxies with $M_{1500}\leq-13$ is $0.86-0.90$ in the three models with constant $\fesc$, but only $0.57$ in their `early heating' model, and the spread at more luminous brackets is similarly dominated by the $\fesc$ prescription rather than by the radiative feedback model alone.

\begin{figure*}
   \centering {
   \includegraphics[width=1\textwidth]{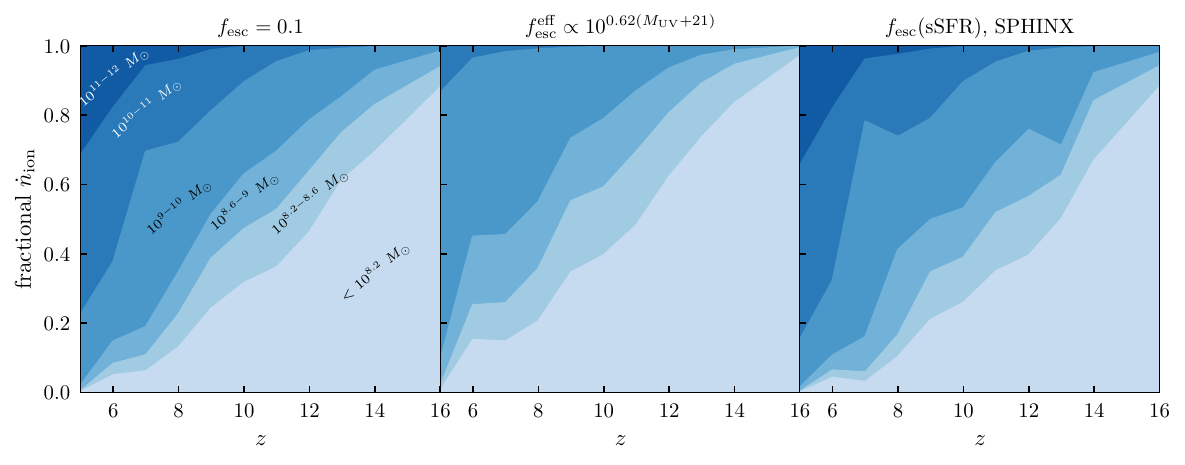}
   }
   \caption{
        Fractional contribution to the total ionizing photon production rate $\dot{n}_{\rm ion}$ from \GRUMPY galaxies of different halo masses as a function of redshift. The panels show the three escape fraction models considered in \S~\ref{subsec:fesc_models}. Shading from light to dark corresponds to increasing halo mass labeled in the left panel. Fractions are normalized to the total $\dot{n}_{\rm ion}$ of all model
        galaxies. The mass bins follow those of \citet[][Fig. C1]{Hutter.etal.2021} for comparison, though see also Fig. 14 of \citet{Rosdahl.etal.2022}. Our model agree to these studies quite well.
    }
   \label{fig:fracnion_Mh}
\end{figure*}

Figure~\ref{fig:fracnion_Mh} presents the corresponding decomposition by host halo mass for each of the three $\fesc$ models in this study, using similar mass bins to \citet[][their Fig. C1]{Hutter.etal.2021}. We note that for a constant $\fesc$ the fractional decomposition is independent of the assumed value, so the left panel equivalently represents the ionizing emissivity from galaxies. Comparing the panels shows that the choice of $\fesc$ model shifts the halo mass scale that dominates the ionizing photon budget: relative to the constant $\fesc$ case, the SPHINX-motivated $\fesc({\rm sSFR})$ model produces a similar result, though the strongly luminosity-dependent model moves the bulk of the contribution to halos of lower mass bins.
Note that our lowest mass bin, $M_{\rm h}<10^{8.2}\,M_\odot$, has no counterpart in their simulation, which does not resolve halos below this mass. Excluding this bin and re-normalizing the remaining fractions, the fractions and the trends with halo mass in our model agree well with the results of \citet{Hutter.etal.2021}. These trends are also qualitatively similar to the corresponding decomposition in SPHINX simulations presented in Fig. 14 of \citealt{Rosdahl.etal.2022}.

These differences in the halo mass decomposition reflect the dependence of $\fesc$ on halo masses of each model. Recent studies have shown or adopted $\fesc$ that decreases with higher halo mass: in the `early heating' model of \citet{Hutter.etal.2021}, $\fesc$ decreases from $\approx0.6$ in halos of $M_{\rm h}\lesssim10^{9}\,M_\odot$ to $\sim10^{-2}$ at $M_{\rm h}\gtrsim10^{11}\,M_\odot$ (their Fig. 5). \citet{Ocvirk.etal.2021} and \citet{Lewis.etal.2023} find $\fesc\approx1$ in halos with $M_{\rm h}\lesssim10^{8.5}\,M_\odot$, decreasing by $2 \sim 3$ orders of magnitude by $10^{11}\,M_\odot$, with a steeper decline when the star formation criterion is more stringent. 

\citet{Mutch.etal.2024} calibrated a sigmoid function dependent on $M_{\rm vir}$ against observational constraints, and found a transition from $\fesc\approx0.16$ to $0.05$ at $M_{\rm vir}\approx10^{9.3}\,M_\odot$. The median $\fesc$ in our $\fesc({\rm sSFR})$ model is nearly constant at $\fesc\approx 1-5\times 10^-3$ at $7.5\lesssim \log_{10}M_{\rm vir}/M_\odot\lesssim 12$ and drops by a factor of $\sim 5-6$ at larger masses. than an order of magnitude over the range of $10^{7.5}< M_{\rm h} < 10^{12}\,M_\odot$. This is similar to the behavior of the escape fractions in the SPHINX simulations on which our model is based, especially at higher redshifts \citep[][see their Fig. 8]{Rosdahl.etal.2022}. This weak mass dependence can also be observed in the fractional $\dnion$ decomposition by halo mass in Fig.~\ref{fig:fracnion_Mh}, where the $\fesc({\rm sSFR})$ model is similar to that of the constant 10\% $\fesc$ model (panels 1 and 3). Despite the low median values of $\fesc$ and weak trend with halo mass in our model, the $\fesc$ values for individual galaxies in our model have large scatter due to a large scatter in the sSFR$_{\rm max,10}$, as star formation rates in high-redshift galaxies in our model are quite bursty (see Fig.~\ref{fig:SPHINX_sSFR}). Thus, at all masses at a given epoch there are galaxies that have high escape fraction and that contribute significantly to the ionizing photon budget. 

It is worth noting that all of the aforementioned models produce reionization histories consistent with observational constraints, despite substantial differences in fractional contributions from different galaxies to the ionizing photon budget. This indicates that current constraints on $Q_{\rm HII}(z)$ and $\tau$ primarily constrain the integrated ionizing photon emissivity, rather than the galactic escape fraction of galaxies of different mass, consistent with our conclusion that $\fesc$ is the main source of uncertainty in modeling reionization.

\section{Summary and conclusions}
\label{sec:summary}

We presented model calculations of the reionization history of hydrogen using a galaxy formation model, which reproduces observational monochromatic ($\lambda=1500\,\aa$) UV luminosity function measurements at $z\in [5,16]$ and basic properties of both local dwarf galaxies \citep[][]{Manwadkar.Kravtsov.2022} and high-redshift galaxies \citep[][]{Kravtsov.Belokurov.2024, Wu.Kravtsov.2024}. The star formation histories of model galaxies, including accounting for stochastic/bursty mode of star formation, have been used to estimate the Lyman continuum photon flux density function for galaxies at redshifts $z\in[5,16]$ including the contribution of galaxies over the broad UV luminosity range from $\Muv\approx -25$ to $\Muv=-5$.\footnote{The best fit parameters of the modified Schechter function describing the ionizing flux density of galaxies as a function of UV luminosity are provided in Table \ref{tab:nion_params} in the Appendix~\ref{app:nion_params}.} 

We use the ionizing photon density functions along with different models for the escape fraction of ionizing photons, $\fesc$, to study the effects of accounting galaxies of different UV luminosities and different assumptions about $\fesc$ on the reionization history of the Universe, quantified by the hydrogen ionized fraction $Q_{\rm HII}(z)$ and its complementary neutral hydrogen fraction. 
Our main results and conclusions are as follows:

\begin{itemize}

\item[1.] We show that including progressively fainter galaxies accelerates reionization (Figure~\ref{fig:Qz_muv}). Accounting for the contribution of faint galaxies with UV luminosities $\Muv>-13$ in additional to bright galaxies results in the hydrogen reionization history, $Q_{\rm HII}(z)$, consistent with all current observational constraints for the {\it constant} ionizing photon escape fraction as low as $\fesc=0.1$ (Figures~\ref{fig:Qz_fesc}-\ref{fig:tauz_fesc}).

\item[2.] Comparing results of the constant $\fesc$ model and two alternative models for $\fesc$ dependence on galaxy properties, motivated by results of different galaxy formation simulations,  shows high sensitivity of $Q_{\rm HII}(z)$ to the assumptions about the escape fraction behavior in the faint UV luminosity regime (Figures~\ref{fig:log1Qz_fesc}-\ref{fig:Qz_fesc}). 

\item[3.] The model with strong luminosity dependence of $\fesc$, which assigns high escape fractions to faint galaxies, results in early hydrogen reionization inconsistent with observational constraints on $Q_{\rm HII}(z)$ and the optical depth for the Thomson scattering of CMB photons, which is in agreement with conclusions of \citet{Munoz.etal.2024}. 

\item[4.] The model in which $\fesc$ follows a universal redshift-independent correlation with the recent maximum specific star formation rate, $\rm sSFR_{10,\rm max}$, qualitatively similar to the correlation found in the SPHINX galaxy formation simulation, results in the reionization history in good agreement with existing observational constraints. Interestingly, this model produces a sizeable ionized hydrogen fraction of $Q_{\rm HII}\approx 0.15-0.2$ at redshifts $z=8-12$ (Figure~\ref{fig:Qz_fesc}). 
\end{itemize}

Our results show that although dwarf galaxies beyond the range of luminosities probed by HST and JWST produce a significant fraction of total hydrogen ionizing photons \citep[as shown in our previous study][see also Figure~\ref{fig:Qz_muv} above]{Wu.Kravtsov.2024}, their relative contribution depends sensitively on the assumptions about the escape fraction of ionizing photons for galaxies of different luminosity. The escape fraction of ionizing Lyman continuum photons and its dependence on galaxy properties at high redshifts is thus the main source of uncertainty in modeling details of hydrogen reionization, and should be the main focus of observational studies and galaxy formation simulations.

\section*{Acknowledgements}

We are grateful to Brant Robertson, Nicholas Choustikov, Piero Madau, and the UChicago structure formation group for useful discussions. We also thank Nickolay Gnedin for discussions and useful detailed comments on the draft of this paper. 
ZW was supported by the University of Chicago CCRF's Quad Research Scholarship, University of California San Diego's Astronomy \& Astrophysics Achievement Award. ZW is grateful to the University of Chicago for supporting his undergraduate thesis, which formed the basis for this work. The author thanks Hsiao-Wen Chen for valuable feedback as a member of his undergraduate thesis committee. ZW also thanks the LSST-DA Data Science Fellowship Program, which is funded by LSST-DA, the Brinson Foundation, the WoodNext Foundation, the Research Corporation for Science Advancement Foundation, and Kevin Wells; His participation in the program has benefited this work.
AK was supported via the National Science Foundation grants AST-1911111 and AST-2408267, and NASA ATP grant 80NSSC20K0512. 

Analyses presented in this paper were greatly aided by the following free software packages: {\tt NumPy} \citep{NumPy}, {\tt SciPy} \citep{scipy}, {\tt Matplotlib} \citep{matplotlib}, {\tt FSPS} \citep{fsps} and its Python bindings package {\tt Python-FSPS}\footnote{\href{https://github.com/dfm/python-fsps}{\tt https://github.com/dfm/python-fsps}}, {\tt BPASS} stellar population synthesis tables for ionizing luminosity \cite{Byrne.etal.2022}, and {\tt Colossus} cosmology package \citep{colossus}. We have also used the Astrophysics Data Service (\href{http://adsabs.harvard.edu/abstract_service.html}{\tt ADS}) and \href{https://arxiv.org}{\tt arXiv} preprint repository extensively during this project and the writing of the paper.

\section*{Data Availability}

A \GRUMPY model implementation is available at \url{https://github.com/virajvman/GRUMPY}. Results of this paper can be reproduced from the fitting functions and parameters presented in Appendices~\ref{app:uvlf_fit} and \ref{app:nion_params}. The data used in the plots within this article are available on request to the corresponding author.

\clearpage
\bibliographystyle{mnras}
\bibliography{reionization}

\begin{thebibliography}{}
\makeatletter
\relax
\def\mn@urlcharsother{\let\do\@makeother \do\$\do\&\do\#\do\^\do\_\do\%\do\~}
\def\mn@doi{\begingroup\mn@urlcharsother \@ifnextchar [ {\mn@doi@} {\mn@doi@[]}}
\def\mn@doi@[#1]#2{\def\@tempa{#1}\ifx\@tempa\@empty \href {http://dx.doi.org/#2} {doi:#2}\else \href {http://dx.doi.org/#2} {#1}\fi \endgroup}
\def\mn@eprint#1#2{\mn@eprint@#1:#2::\@nil}
\def\mn@eprint@arXiv#1{\href {http://arxiv.org/abs/#1} {{\tt arXiv:#1}}}
\def\mn@eprint@dblp#1{\href {http://dblp.uni-trier.de/rec/bibtex/#1.xml} {dblp:#1}}
\def\mn@eprint@#1:#2:#3:#4\@nil{\def\@tempa {#1}\def\@tempb {#2}\def\@tempc {#3}\ifx \@tempc \@empty \let \@tempc \@tempb \let \@tempb \@tempa \fi \ifx \@tempb \@empty \def\@tempb {arXiv}\fi \@ifundefined {mn@eprint@\@tempb}{\@tempb:\@tempc}{\expandafter \expandafter \csname mn@eprint@\@tempb\endcsname \expandafter{\@tempc}}}

\bibitem[\protect\citeauthoryear{{Acharya} et~al.,}{{Acharya} et~al.}{2025}]{Acharya.etal.2025}
{Acharya} A.,  et~al., 2025, \mn@doi [\mnras] {10.1093/mnras/staf1412}, \href {https://ui.adsabs.harvard.edu/abs/2025MNRAS.543.1058A} {543, 1058}

\bibitem[\protect\citeauthoryear{{Anderson}, {Governato}, {Karcher}, {Quinn}  \& {Wadsley}}{{Anderson} et~al.}{2017}]{Anderson.etal.2017}
{Anderson} L.,  {Governato} F.,  {Karcher} M.,  {Quinn} T.,   {Wadsley} J.,  2017, \mn@doi [\mnras] {10.1093/mnras/stx709}, \href {https://ui.adsabs.harvard.edu/abs/2017MNRAS.468.4077A} {468, 4077}

\bibitem[\protect\citeauthoryear{{Becker} et~al.,}{{Becker} et~al.}{2001}]{Becker.etal.2001}
{Becker} R.~H.,  et~al., 2001, \mn@doi [\aj] {10.1086/324231}, \href {https://ui.adsabs.harvard.edu/abs/2001AJ....122.2850B} {122, 2850}

\bibitem[\protect\citeauthoryear{{Bosman} et~al.,}{{Bosman} et~al.}{2022}]{Bosman.etal.2022}
{Bosman} S. E.~I.,  et~al., 2022, \mn@doi [\mnras] {10.1093/mnras/stac1046}, \href {https://ui.adsabs.harvard.edu/abs/2022MNRAS.514...55B} {514, 55}

\bibitem[\protect\citeauthoryear{{Bouwens} et~al.,}{{Bouwens} et~al.}{2021}]{Bouwens.etal.2021}
{Bouwens} R.~J.,  et~al., 2021, \mn@doi [\aj] {10.3847/1538-3881/abf83e}, \href {https://ui.adsabs.harvard.edu/abs/2021AJ....162...47B} {162, 47}

\bibitem[\protect\citeauthoryear{{Bouwens}, {Illingworth}, {Ellis}, {Oesch}  \& {Stefanon}}{{Bouwens} et~al.}{2022}]{Bouwens.etal.2022}
{Bouwens} R.~J.,  {Illingworth} G.,  {Ellis} R.~S.,  {Oesch} P.,   {Stefanon} M.,  2022, \mn@doi [\apj] {10.3847/1538-4357/ac86d1}, \href {https://ui.adsabs.harvard.edu/abs/2022ApJ...940...55B} {940, 55}

\bibitem[\protect\citeauthoryear{{Bouwens}, {Illingworth}, {Oesch}, {Stefanon}, {Naidu}, {van Leeuwen}  \& {Magee}}{{Bouwens} et~al.}{2023a}]{Bouwens.etal.2023a}
{Bouwens} R.,  {Illingworth} G.,  {Oesch} P.,  {Stefanon} M.,  {Naidu} R.,  {van Leeuwen} I.,   {Magee} D.,  2023a, \mn@doi [\mnras] {10.1093/mnras/stad1014}, \href {https://ui.adsabs.harvard.edu/abs/2023MNRAS.523.1009B} {523, 1009}

\bibitem[\protect\citeauthoryear{{Bouwens} et~al.,}{{Bouwens} et~al.}{2023b}]{Bouwens.etal.2023b}
{Bouwens} R.~J.,  et~al., 2023b, \mn@doi [\mnras] {10.1093/mnras/stad1145}, \href {https://ui.adsabs.harvard.edu/abs/2023MNRAS.523.1036B} {523, 1036}

\bibitem[\protect\citeauthoryear{{Boylan-Kolchin}, {Weisz}, {Johnson}, {Bullock}, {Conroy}  \& {Fitts}}{{Boylan-Kolchin} et~al.}{2015}]{BoylanKolchin.etal.2015}
{Boylan-Kolchin} M.,  {Weisz} D.~R.,  {Johnson} B.~D.,  {Bullock} J.~S.,  {Conroy} C.,   {Fitts} A.,  2015, \mn@doi [\mnras] {10.1093/mnras/stv1736}, \href {https://ui.adsabs.harvard.edu/abs/2015MNRAS.453.1503B} {453, 1503}

\bibitem[\protect\citeauthoryear{{Byrne}, {Stanway}, {Eldridge}, {McSwiney}  \& {Townsend}}{{Byrne} et~al.}{2022}]{Byrne.etal.2022}
{Byrne} C.~M.,  {Stanway} E.~R.,  {Eldridge} J.~J.,  {McSwiney} L.,   {Townsend} O.~T.,  2022, \mn@doi [\mnras] {10.1093/mnras/stac807}, \href {https://ui.adsabs.harvard.edu/abs/2022MNRAS.512.5329B} {512, 5329}

\bibitem[\protect\citeauthoryear{{Caplar} \& {Tacchella}}{{Caplar} \& {Tacchella}}{2019}]{Caplar.Tacchella.2019}
{Caplar} N.,  {Tacchella} S.,  2019, \mn@doi [\mnras] {10.1093/mnras/stz1449}, \href {https://ui.adsabs.harvard.edu/abs/2019MNRAS.487.3845C} {487, 3845}

\bibitem[\protect\citeauthoryear{{Chemerynska} et~al.,}{{Chemerynska} et~al.}{2024}]{Chemerynska.etal.2024}
{Chemerynska} I.,  et~al., 2024, \mn@doi [\mnras] {10.1093/mnras/stae1260}, \href {https://ui.adsabs.harvard.edu/abs/2024MNRAS.531.2615C} {531, 2615}

\bibitem[\protect\citeauthoryear{{Chemerynska} et~al.,}{{Chemerynska} et~al.}{2026}]{Chemerynska.etal.2026}
{Chemerynska} I.,  et~al., 2026, \mn@doi [\mnras] {10.1093/mnras/staf2267}, \href {https://ui.adsabs.harvard.edu/abs/2026MNRAS.546f2267C} {546, staf2267}

\bibitem[\protect\citeauthoryear{{Chen}, {Doussot}, {Trac}  \& {Cen}}{{Chen} et~al.}{2020}]{Chen.etal.2020}
{Chen} N.,  {Doussot} A.,  {Trac} H.,   {Cen} R.,  2020, \mn@doi [\apj] {10.3847/1538-4357/abc890}, \href {https://ui.adsabs.harvard.edu/abs/2020ApJ...905..132C} {905, 132}

\bibitem[\protect\citeauthoryear{{Chisholm} et~al.,}{{Chisholm} et~al.}{2022}]{Chisholm.etal.2022}
{Chisholm} J.,  et~al., 2022, \mn@doi [\mnras] {10.1093/mnras/stac2874}, \href {https://ui.adsabs.harvard.edu/abs/2022MNRAS.517.5104C} {517, 5104}

\bibitem[\protect\citeauthoryear{{Choustikov} et~al.,}{{Choustikov} et~al.}{2024a}]{Choustikov.etal.2024}
{Choustikov} N.,  et~al., 2024a, \mn@doi [\mnras] {10.1093/mnras/stae776}, \href {https://ui.adsabs.harvard.edu/abs/2024MNRAS.529.3751C} {529, 3751}

\bibitem[\protect\citeauthoryear{{Choustikov} et~al.,}{{Choustikov} et~al.}{2024b}]{Choustikov.etal.2024b}
{Choustikov} N.,  et~al., 2024b, \mn@doi [\mnras] {10.1093/mnras/stae1586}, \href {https://ui.adsabs.harvard.edu/abs/2024MNRAS.532.2463C} {532, 2463}

\bibitem[\protect\citeauthoryear{{Choustikov}, {Stiskalek}, {Saxena}, {Katz}, {Devriendt}, {Slyz}  \& {Sub-department of Astrophysics}}{{Choustikov} et~al.}{2025}]{Choustikov.etal.2025}
{Choustikov} N.,  {Stiskalek} R.,  {Saxena} A.,  {Katz} H.,  {Devriendt} J.,  {Slyz} A.,   {Sub-department of Astrophysics} U. o.~O.,  2025, \mn@doi [\mnras] {10.1093/mnras/staf126}, \href {https://ui.adsabs.harvard.edu/abs/2025MNRAS.537.2273C} {537, 2273}

\bibitem[\protect\citeauthoryear{{Conroy} \& {Gunn}}{{Conroy} \& {Gunn}}{2010a}]{fsps}
{Conroy} C.,  {Gunn} J.~E.,  2010a, {FSPS: Flexible Stellar Population Synthesis}, Astrophysics Source Code Library, record ascl:1010.043

\bibitem[\protect\citeauthoryear{{Conroy} \& {Gunn}}{{Conroy} \& {Gunn}}{2010b}]{Conroy.Gunn.2010}
{Conroy} C.,  {Gunn} J.~E.,  2010b, \mn@doi [\apj] {10.1088/0004-637X/712/2/833}, \href {https://ui.adsabs.harvard.edu/abs/2010ApJ...712..833C} {712, 833}

\bibitem[\protect\citeauthoryear{{Conroy}, {Gunn}  \& {White}}{{Conroy} et~al.}{2009}]{Conroy.etal.2009}
{Conroy} C.,  {Gunn} J.~E.,   {White} M.,  2009, \mn@doi [\apj] {10.1088/0004-637X/699/1/486}, \href {https://ui.adsabs.harvard.edu/abs/2009ApJ...699..486C} {699, 486}

\bibitem[\protect\citeauthoryear{{Cullen} et~al.,}{{Cullen} et~al.}{2023}]{Cullen.etal.2023}
{Cullen} F.,  et~al., 2023, \mn@doi [\mnras] {10.1093/mnras/stad073}, \href {https://ui.adsabs.harvard.edu/abs/2023MNRAS.520...14C} {520, 14}

\bibitem[\protect\citeauthoryear{{Davies} et~al.,}{{Davies} et~al.}{2018}]{Davies.etal.2018}
{Davies} F.~B.,  et~al., 2018, \mn@doi [\apj] {10.3847/1538-4357/aad6dc}, \href {https://ui.adsabs.harvard.edu/abs/2018ApJ...864..142D} {864, 142}

\bibitem[\protect\citeauthoryear{{Davies}, {Bosman}  \& {Furlanetto}}{{Davies} et~al.}{2024a}]{Davies.etal.2024b}
{Davies} F.~B.,  {Bosman} S. E.~I.,   {Furlanetto} S.~R.,  2024a, \mn@doi [arXiv e-prints] {10.48550/arXiv.2406.18186}, \href {https://ui.adsabs.harvard.edu/abs/2024arXiv240618186D} {p. arXiv:2406.18186}

\bibitem[\protect\citeauthoryear{{Davies} et~al.,}{{Davies} et~al.}{2024b}]{Davies.etal.2024}
{Davies} F.~B.,  et~al., 2024b, \mn@doi [\apj] {10.3847/1538-4357/ad1d5d}, \href {https://ui.adsabs.harvard.edu/abs/2024ApJ...965..134D} {965, 134}

\bibitem[\protect\citeauthoryear{{Davies} et~al.,}{{Davies} et~al.}{2025}]{Davies.etal.2025}
{Davies} F.~B.,  et~al., 2025, \mn@doi [\mnras] {10.1093/mnras/staf1862}, \href {https://ui.adsabs.harvard.edu/abs/2025MNRAS.tmp.1755D} {}

\bibitem[\protect\citeauthoryear{{Diemer}}{{Diemer}}{2018}]{colossus}
{Diemer} B.,  2018, \mn@doi [\apjs] {10.3847/1538-4365/aaee8c}, \href {https://ui.adsabs.harvard.edu/abs/2018ApJS..239...35D} {239, 35}

\bibitem[\protect\citeauthoryear{{Donnan} et~al.,}{{Donnan} et~al.}{2023}]{Donnan.etal.2023}
{Donnan} C.~T.,  et~al., 2023, \mn@doi [\mnras] {10.1093/mnras/stac3472}, \href {https://ui.adsabs.harvard.edu/abs/2023MNRAS.518.6011D} {518, 6011}

\bibitem[\protect\citeauthoryear{{Donnan} et~al.,}{{Donnan} et~al.}{2024}]{Donnan.etal.2024}
{Donnan} C.~T.,  et~al., 2024, \mn@doi [\mnras] {10.1093/mnras/stae2037}, \href {https://ui.adsabs.harvard.edu/abs/2024MNRAS.533.3222D} {533, 3222}

\bibitem[\protect\citeauthoryear{{Faucher-Gigu{\`e}re}}{{Faucher-Gigu{\`e}re}}{2020}]{Faucher-Giguere.2020}
{Faucher-Gigu{\`e}re} C.-A.,  2020, \mn@doi [\mnras] {10.1093/mnras/staa302}, \href {https://ui.adsabs.harvard.edu/abs/2020MNRAS.493.1614F} {493, 1614}

\bibitem[\protect\citeauthoryear{{Ferrara}, {Giavalisco}, {Pentericci}, {Vanzella}, {Calabr{\`o}}  \& {Llerena}}{{Ferrara} et~al.}{2025}]{Ferrara.etal.2025}
{Ferrara} A.,  {Giavalisco} M.,  {Pentericci} L.,  {Vanzella} E.,  {Calabr{\`o}} A.,   {Llerena} M.,  2025, \mn@doi [The Open Journal of Astrophysics] {10.33232/001c.143600}, \href {https://ui.adsabs.harvard.edu/abs/2025OJAp....8E.125F} {8, 125}

\bibitem[\protect\citeauthoryear{{Finkelstein} et~al.,}{{Finkelstein} et~al.}{2019}]{Finkelstein.etal.2019}
{Finkelstein} S.~L.,  et~al., 2019, \mn@doi [\apj] {10.3847/1538-4357/ab1ea8}, \href {https://ui.adsabs.harvard.edu/abs/2019ApJ...879...36F} {879, 36}

\bibitem[\protect\citeauthoryear{{Finkelstein} et~al.,}{{Finkelstein} et~al.}{2022}]{Finkelstein.etal.2022}
{Finkelstein} S.~L.,  et~al., 2022, \mn@doi [\apjl] {10.3847/2041-8213/ac966e}, \href {https://ui.adsabs.harvard.edu/abs/2022ApJ...940L..55F} {940, L55}

\bibitem[\protect\citeauthoryear{{Finkelstein} et~al.,}{{Finkelstein} et~al.}{2023}]{Finkelstein.etal.2023}
{Finkelstein} S.~L.,  et~al., 2023, \mn@doi [\apjl] {10.3847/2041-8213/acade4}, \href {https://ui.adsabs.harvard.edu/abs/2023ApJ...946L..13F} {946, L13}

\bibitem[\protect\citeauthoryear{{Finlator}, {{\"O}zel}  \& {Dav{\'e}}}{{Finlator} et~al.}{2009}]{Finlator.etal.2009}
{Finlator} K.,  {{\"O}zel} F.,   {Dav{\'e}} R.,  2009, \mn@doi [\mnras] {10.1111/j.1365-2966.2008.14190.x}, \href {https://ui.adsabs.harvard.edu/abs/2009MNRAS.393.1090F} {393, 1090}

\bibitem[\protect\citeauthoryear{{Finlator}, {Oh}, {{\"O}zel}  \& {Dav{\'e}}}{{Finlator} et~al.}{2012}]{Finlator.etal.2012}
{Finlator} K.,  {Oh} S.~P.,  {{\"O}zel} F.,   {Dav{\'e}} R.,  2012, \mn@doi [\mnras] {10.1111/j.1365-2966.2012.22114.x}, \href {https://ui.adsabs.harvard.edu/abs/2012MNRAS.427.2464F} {427, 2464}

\bibitem[\protect\citeauthoryear{{Franco} et~al.,}{{Franco} et~al.}{2025}]{Franco.etal.2025}
{Franco} M.,  et~al., 2025, \mn@doi [arXiv e-prints] {10.48550/arXiv.2508.04791}, \href {https://ui.adsabs.harvard.edu/abs/2025arXiv250804791F} {p. arXiv:2508.04791}

\bibitem[\protect\citeauthoryear{{Gaikwad} et~al.,}{{Gaikwad} et~al.}{2023}]{Gaikwad.etal.2023}
{Gaikwad} P.,  et~al., 2023, \mn@doi [\mnras] {10.1093/mnras/stad2566}, \href {https://ui.adsabs.harvard.edu/abs/2023MNRAS.525.4093G} {525, 4093}

\bibitem[\protect\citeauthoryear{{Gnedin}}{{Gnedin}}{2000}]{Gnedin.2000a}
{Gnedin} N.~Y.,  2000, \mn@doi [\apj] {10.1086/308876}, \href {https://ui.adsabs.harvard.edu/abs/2000ApJ...535..530G} {535, 530}

\bibitem[\protect\citeauthoryear{{Gnedin}}{{Gnedin}}{2004}]{Gnedin.2004}
{Gnedin} N.~Y.,  2004, \mn@doi [\apj] {10.1086/421450}, \href {https://ui.adsabs.harvard.edu/abs/2004ApJ...610....9G} {610, 9}

\bibitem[\protect\citeauthoryear{{Gnedin}}{{Gnedin}}{2016}]{Gnedin.2016}
{Gnedin} N.~Y.,  2016, \mn@doi [\apjl] {10.3847/2041-8205/825/2/L17}, \href {https://ui.adsabs.harvard.edu/abs/2016ApJ...825L..17G} {825, L17}

\bibitem[\protect\citeauthoryear{{Gnedin}}{{Gnedin}}{2024}]{Gnedin.2024}
{Gnedin} N.~Y.,  2024, \mn@doi [\apj] {10.3847/1538-4357/ad298e}, \href {https://ui.adsabs.harvard.edu/abs/2024ApJ...963..150G} {963, 150}

\bibitem[\protect\citeauthoryear{{Gnedin} \& {Fan}}{{Gnedin} \& {Fan}}{2006}]{Gnedin.Fan.2006}
{Gnedin} N.~Y.,  {Fan} X.,  2006, \mn@doi [\apj] {10.1086/505790}, \href {https://ui.adsabs.harvard.edu/abs/2006ApJ...648....1G} {648, 1}

\bibitem[\protect\citeauthoryear{{Gnedin} \& {Kaurov}}{{Gnedin} \& {Kaurov}}{2014}]{Gnedin.Kaurov.2014}
{Gnedin} N.~Y.,  {Kaurov} A.~A.,  2014, \mn@doi [\apj] {10.1088/0004-637X/793/1/30}, \href {https://ui.adsabs.harvard.edu/abs/2014ApJ...793...30G} {793, 30}

\bibitem[\protect\citeauthoryear{{Gnedin} \& {Madau}}{{Gnedin} \& {Madau}}{2022}]{Gnedin.Madau.2022}
{Gnedin} N.~Y.,  {Madau} P.,  2022, \mn@doi [Living Reviews in Computational Astrophysics] {10.1007/s41115-022-00015-5}, \href {https://ui.adsabs.harvard.edu/abs/2022LRCA....8....3G} {8, 3}

\bibitem[\protect\citeauthoryear{{Gnedin}, {Kravtsov}  \& {Chen}}{{Gnedin} et~al.}{2008}]{Gnedin.etal.2008}
{Gnedin} N.~Y.,  {Kravtsov} A.~V.,   {Chen} H.-W.,  2008, \mn@doi [\apj] {10.1086/524007}, \href {https://ui.adsabs.harvard.edu/abs/2008ApJ...672..765G} {672, 765}

\bibitem[\protect\citeauthoryear{{Greig}, {Mesinger}, {Haiman}  \& {Simcoe}}{{Greig} et~al.}{2017}]{Greig.etal.2017}
{Greig} B.,  {Mesinger} A.,  {Haiman} Z.,   {Simcoe} R.~A.,  2017, \mn@doi [\mnras] {10.1093/mnras/stw3351}, \href {https://ui.adsabs.harvard.edu/abs/2017MNRAS.466.4239G} {466, 4239}

\bibitem[\protect\citeauthoryear{{Gunn} \& {Peterson}}{{Gunn} \& {Peterson}}{1965}]{Gunn.Peterson.1965}
{Gunn} J.~E.,  {Peterson} B.~A.,  1965, \mn@doi [\apj] {10.1086/148444}, \href {https://ui.adsabs.harvard.edu/abs/1965ApJ...142.1633G} {142, 1633}

\bibitem[\protect\citeauthoryear{{Harikane} et~al.,}{{Harikane} et~al.}{2023}]{Harikane.etal.2023}
{Harikane} Y.,  et~al., 2023, \mn@doi [\apjs] {10.3847/1538-4365/acaaa9}, \href {https://ui.adsabs.harvard.edu/abs/2023ApJS..265....5H} {265, 5}

\bibitem[\protect\citeauthoryear{{Harikane}, {Nakajima}, {Ouchi}, {Umeda}, {Isobe}, {Ono}, {Xu}  \& {Zhang}}{{Harikane} et~al.}{2024}]{Harikane.etal.2024}
{Harikane} Y.,  {Nakajima} K.,  {Ouchi} M.,  {Umeda} H.,  {Isobe} Y.,  {Ono} Y.,  {Xu} Y.,   {Zhang} Y.,  2024, \mn@doi [\apj] {10.3847/1538-4357/ad0b7e}, \href {https://ui.adsabs.harvard.edu/abs/2024ApJ...960...56H} {960, 56}

\bibitem[\protect\citeauthoryear{Harris et~al.,}{Harris et~al.}{2020}]{NumPy}
Harris C.~R.,  et~al., 2020, \mn@doi [Nature] {10.1038/s41586-020-2649-2}, 585, 357–362

\bibitem[\protect\citeauthoryear{{Heckman} et~al.,}{{Heckman} et~al.}{2011}]{Heckman.etal.2011}
{Heckman} T.~M.,  et~al., 2011, \mn@doi [\apj] {10.1088/0004-637X/730/1/5}, \href {https://ui.adsabs.harvard.edu/abs/2011ApJ...730....5H} {730, 5}

\bibitem[\protect\citeauthoryear{{Heinrich} \& {Hu}}{{Heinrich} \& {Hu}}{2021}]{Heinrich.Hu.2021}
{Heinrich} C.,  {Hu} W.,  2021, \mn@doi [\prd] {10.1103/PhysRevD.104.063505}, \href {https://ui.adsabs.harvard.edu/abs/2021PhRvD.104f3505H} {104, 063505}

\bibitem[\protect\citeauthoryear{{Howard}, {Pudritz}  \& {Klessen}}{{Howard} et~al.}{2017}]{Howard.etal.2017}
{Howard} C.,  {Pudritz} R.,   {Klessen} R.,  2017, \mn@doi [\apj] {10.3847/1538-4357/834/1/40}, \href {https://ui.adsabs.harvard.edu/abs/2017ApJ...834...40H} {834, 40}

\bibitem[\protect\citeauthoryear{{Howard}, {Pudritz}, {Harris}  \& {Klessen}}{{Howard} et~al.}{2018}]{Howard.etal.2018}
{Howard} C.~S.,  {Pudritz} R.~E.,  {Harris} W.~E.,   {Klessen} R.~S.,  2018, \mn@doi [\mnras] {10.1093/mnras/stx3276}, \href {https://ui.adsabs.harvard.edu/abs/2018MNRAS.475.3121H} {475, 3121}

\bibitem[\protect\citeauthoryear{Hunter}{Hunter}{2007}]{matplotlib}
Hunter J.~D.,  2007, \mn@doi [Computing In Science \& Engineering] {10.1109/MCSE.2007.55}, 9, 90

\bibitem[\protect\citeauthoryear{{Hutter}, {Dayal}, {Yepes}, {Gottl{\"o}ber}, {Legrand}  \& {Ucci}}{{Hutter} et~al.}{2021}]{Hutter.etal.2021}
{Hutter} A.,  {Dayal} P.,  {Yepes} G.,  {Gottl{\"o}ber} S.,  {Legrand} L.,   {Ucci} G.,  2021, \mn@doi [\mnras] {10.1093/mnras/stab602}, \href {https://ui.adsabs.harvard.edu/abs/2021MNRAS.503.3698H} {503, 3698}

\bibitem[\protect\citeauthoryear{{Hutter}, {Trebitsch}, {Dayal}, {Gottl{\"o}ber}, {Yepes}  \& {Legrand}}{{Hutter} et~al.}{2023}]{Hutter.etal.2023}
{Hutter} A.,  {Trebitsch} M.,  {Dayal} P.,  {Gottl{\"o}ber} S.,  {Yepes} G.,   {Legrand} L.,  2023, \mn@doi [\mnras] {10.1093/mnras/stad2230}, \href {https://ui.adsabs.harvard.edu/abs/2023MNRAS.524.6124H} {524, 6124}

\bibitem[\protect\citeauthoryear{{Jaacks}, {Thompson}  \& {Nagamine}}{{Jaacks} et~al.}{2013}]{Jaacks.etal.2013}
{Jaacks} J.,  {Thompson} R.,   {Nagamine} K.,  2013, \mn@doi [\apj] {10.1088/0004-637X/766/2/94}, \href {https://ui.adsabs.harvard.edu/abs/2013ApJ...766...94J} {766, 94}

\bibitem[\protect\citeauthoryear{Jones, Oliphant, Peterson  et~al.}{Jones et~al.}{01  }]{scipy}
Jones E.,  Oliphant T.,  Peterson P.,   et~al., 2001--, {SciPy}: Open source scientific tools for {Python}, \url {http://www.scipy.org/}

\bibitem[\protect\citeauthoryear{{Kageura} et~al.,}{{Kageura} et~al.}{2025}]{Kageura.etal.2025}
{Kageura} Y.,  et~al., 2025, \mn@doi [\apjs] {10.3847/1538-4365/adc690}, \href {https://ui.adsabs.harvard.edu/abs/2025ApJS..278...33K} {278, 33}

\bibitem[\protect\citeauthoryear{{Kannan}, {Garaldi}, {Smith}, {Pakmor}, {Springel}, {Vogelsberger}  \& {Hernquist}}{{Kannan} et~al.}{2022}]{Kannan.etal.2022}
{Kannan} R.,  {Garaldi} E.,  {Smith} A.,  {Pakmor} R.,  {Springel} V.,  {Vogelsberger} M.,   {Hernquist} L.,  2022, \mn@doi [\mnras] {10.1093/mnras/stab3710}, \href {https://ui.adsabs.harvard.edu/abs/2022MNRAS.511.4005K} {511, 4005}

\bibitem[\protect\citeauthoryear{{Kaurov} \& {Gnedin}}{{Kaurov} \& {Gnedin}}{2015}]{Kaurov.Gnedin.2015}
{Kaurov} A.~A.,  {Gnedin} N.~Y.,  2015, \mn@doi [\apj] {10.1088/0004-637X/810/2/154}, \href {https://ui.adsabs.harvard.edu/abs/2015ApJ...810..154K} {810, 154}

\bibitem[\protect\citeauthoryear{{Kimm}, {Katz}, {Haehnelt}, {Rosdahl}, {Devriendt}  \& {Slyz}}{{Kimm} et~al.}{2017}]{Kimm.etal.2017}
{Kimm} T.,  {Katz} H.,  {Haehnelt} M.,  {Rosdahl} J.,  {Devriendt} J.,   {Slyz} A.,  2017, \mn@doi [\mnras] {10.1093/mnras/stx052}, \href {https://ui.adsabs.harvard.edu/abs/2017MNRAS.466.4826K} {466, 4826}

\bibitem[\protect\citeauthoryear{{Kravtsov} \& {Belokurov}}{{Kravtsov} \& {Belokurov}}{2024}]{Kravtsov.Belokurov.2024}
{Kravtsov} A.,  {Belokurov} V.,  2024, arXiv/2405.04578, submitted

\bibitem[\protect\citeauthoryear{{Kravtsov} \& {Manwadkar}}{{Kravtsov} \& {Manwadkar}}{2022}]{Kravtsov.Manwadkar.2022}
{Kravtsov} A.,  {Manwadkar} V.,  2022, \mn@doi [\mnras] {10.1093/mnras/stac1439}, \href {https://ui.adsabs.harvard.edu/abs/2022MNRAS.514.2667K} {514, 2667}

\bibitem[\protect\citeauthoryear{{Kravtsov} \& {Wu}}{{Kravtsov} \& {Wu}}{2023}]{Kravtsov.Wu.2023}
{Kravtsov} A.,  {Wu} Z.,  2023, \mn@doi [\mnras] {10.1093/mnras/stad2219}, \href {https://ui.adsabs.harvard.edu/abs/2023MNRAS.525..325K} {525, 325}

\bibitem[\protect\citeauthoryear{{Kuhlen} \& {Faucher-Gigu{\`e}re}}{{Kuhlen} \& {Faucher-Gigu{\`e}re}}{2012}]{Kuhlen.Faucher-Gigu`ere.2012}
{Kuhlen} M.,  {Faucher-Gigu{\`e}re} C.-A.,  2012, \mn@doi [\mnras] {10.1111/j.1365-2966.2012.20924.x}, \href {https://ui.adsabs.harvard.edu/abs/2012MNRAS.423..862K} {423, 862}

\bibitem[\protect\citeauthoryear{{Leung} et~al.,}{{Leung} et~al.}{2023}]{Leung.etal.2023}
{Leung} G. C.~K.,  et~al., 2023, \mn@doi [\apjl] {10.3847/2041-8213/acf365}, \href {https://ui.adsabs.harvard.edu/abs/2023ApJ...954L..46L} {954, L46}

\bibitem[\protect\citeauthoryear{{Lewis}, {Ocvirk}, {Dubois}, {Aubert}, {Chardin}, {Gillet}  \& {Th{\'e}lie}}{{Lewis} et~al.}{2023}]{Lewis.etal.2023}
{Lewis} J. S.~W.,  {Ocvirk} P.,  {Dubois} Y.,  {Aubert} D.,  {Chardin} J.,  {Gillet} N.,   {Th{\'e}lie} {\'E}.,  2023, \mn@doi [\mnras] {10.1093/mnras/stad081}, \href {https://ui.adsabs.harvard.edu/abs/2023MNRAS.519.5987L} {519, 5987}

\bibitem[\protect\citeauthoryear{{Lin} et~al.,}{{Lin} et~al.}{2024}]{Lin.etal.2024}
{Lin} Y.-H.,  et~al., 2024, \mn@doi [\mnras] {10.1093/mnras/stad3483}, \href {https://ui.adsabs.harvard.edu/abs/2024MNRAS.527.4173L} {527, 4173}

\bibitem[\protect\citeauthoryear{{Ma}, {Kasen}, {Hopkins}, {Faucher-Gigu{\`e}re}, {Quataert}, {Kere{\v{s}}}  \& {Murray}}{{Ma} et~al.}{2015}]{Ma.etal.2015}
{Ma} X.,  {Kasen} D.,  {Hopkins} P.~F.,  {Faucher-Gigu{\`e}re} C.-A.,  {Quataert} E.,  {Kere{\v{s}}} D.,   {Murray} N.,  2015, \mn@doi [\mnras] {10.1093/mnras/stv1679}, \href {https://ui.adsabs.harvard.edu/abs/2015MNRAS.453..960M} {453, 960}

\bibitem[\protect\citeauthoryear{{Madau}}{{Madau}}{2017}]{Madau.2017}
{Madau} P.,  2017, \mn@doi [\apj] {10.3847/1538-4357/aa9715}, \href {https://ui.adsabs.harvard.edu/abs/2017ApJ...851...50M} {851, 50}

\bibitem[\protect\citeauthoryear{{Madau}, {Haardt}  \& {Rees}}{{Madau} et~al.}{1999}]{Madau.etal.1999}
{Madau} P.,  {Haardt} F.,   {Rees} M.~J.,  1999, \mn@doi [\apj] {10.1086/306975}, \href {https://ui.adsabs.harvard.edu/abs/1999ApJ...514..648M} {514, 648}

\bibitem[\protect\citeauthoryear{{Madau}, {Giallongo}, {Grazian}  \& {Haardt}}{{Madau} et~al.}{2024}]{Madau.etal.2024}
{Madau} P.,  {Giallongo} E.,  {Grazian} A.,   {Haardt} F.,  2024, \mn@doi [\apj] {10.3847/1538-4357/ad5ce8}, \href {https://ui.adsabs.harvard.edu/abs/2024ApJ...971...75M} {971, 75}

\bibitem[\protect\citeauthoryear{{Manwadkar} \& {Kravtsov}}{{Manwadkar} \& {Kravtsov}}{2022}]{Manwadkar.Kravtsov.2022}
{Manwadkar} V.,  {Kravtsov} A.~V.,  2022, \mn@doi [\mnras] {10.1093/mnras/stac2452}, \href {https://ui.adsabs.harvard.edu/abs/2022MNRAS.516.3944M} {516, 3944}

\bibitem[\protect\citeauthoryear{{Mason}, {Treu}, {Dijkstra}, {Mesinger}, {Trenti}, {Pentericci}, {de Barros}  \& {Vanzella}}{{Mason} et~al.}{2018}]{Mason.etal.2018}
{Mason} C.~A.,  {Treu} T.,  {Dijkstra} M.,  {Mesinger} A.,  {Trenti} M.,  {Pentericci} L.,  {de Barros} S.,   {Vanzella} E.,  2018, \mn@doi [\apj] {10.3847/1538-4357/aab0a7}, \href {https://ui.adsabs.harvard.edu/abs/2018ApJ...856....2M} {856, 2}

\bibitem[\protect\citeauthoryear{{Mason}, {Naidu}, {Tacchella}  \& {Leja}}{{Mason} et~al.}{2019}]{Mason.etal.2019}
{Mason} C.~A.,  {Naidu} R.~P.,  {Tacchella} S.,   {Leja} J.,  2019, \mn@doi [\mnras] {10.1093/mnras/stz2291}, \href {https://ui.adsabs.harvard.edu/abs/2019MNRAS.489.2669M} {489, 2669}

\bibitem[\protect\citeauthoryear{{McLeod} et~al.,}{{McLeod} et~al.}{2024}]{McLeod.etal.2024}
{McLeod} D.~J.,  et~al., 2024, \mn@doi [\mnras] {10.1093/mnras/stad3471}, \href {https://ui.adsabs.harvard.edu/abs/2024MNRAS.527.5004M} {527, 5004}

\bibitem[\protect\citeauthoryear{{Menon}, {Burkhart}, {Somerville}, {Thompson}  \& {Sternberg}}{{Menon} et~al.}{2025}]{Menon.etal.2025}
{Menon} S.~H.,  {Burkhart} B.,  {Somerville} R.~S.,  {Thompson} T.~A.,   {Sternberg} A.,  2025, \mn@doi [\apj] {10.3847/1538-4357/add2f9}, \href {https://ui.adsabs.harvard.edu/abs/2025ApJ...987...12M} {987, 12}

\bibitem[\protect\citeauthoryear{{Mu{\~n}oz}, {Mirocha}, {Chisholm}, {Furlanetto}  \& {Mason}}{{Mu{\~n}oz} et~al.}{2024}]{Munoz.etal.2024}
{Mu{\~n}oz} J.~B.,  {Mirocha} J.,  {Chisholm} J.,  {Furlanetto} S.~R.,   {Mason} C.,  2024, \mn@doi [\mnras] {10.1093/mnrasl/slae086}, \href {https://ui.adsabs.harvard.edu/abs/2024MNRAS.535L..37M} {535, L37}

\bibitem[\protect\citeauthoryear{{Mutch}, {Greig}, {Qin}, {Poole}  \& {Wyithe}}{{Mutch} et~al.}{2024}]{Mutch.etal.2024}
{Mutch} S.~J.,  {Greig} B.,  {Qin} Y.,  {Poole} G.~B.,   {Wyithe} J. S.~B.,  2024, \mn@doi [\mnras] {10.1093/mnras/stad3746}, \href {https://ui.adsabs.harvard.edu/abs/2024MNRAS.527.7924M} {527, 7924}

\bibitem[\protect\citeauthoryear{{Naidu}, {Tacchella}, {Mason}, {Bose}, {Oesch}  \& {Conroy}}{{Naidu} et~al.}{2020}]{Naidu.etal.2020}
{Naidu} R.~P.,  {Tacchella} S.,  {Mason} C.~A.,  {Bose} S.,  {Oesch} P.~A.,   {Conroy} C.,  2020, \mn@doi [\apj] {10.3847/1538-4357/ab7cc9}, \href {https://ui.adsabs.harvard.edu/abs/2020ApJ...892..109N} {892, 109}

\bibitem[\protect\citeauthoryear{{O'Shea}, {Wise}, {Xu}  \& {Norman}}{{O'Shea} et~al.}{2015}]{OShea.etal.2015}
{O'Shea} B.~W.,  {Wise} J.~H.,  {Xu} H.,   {Norman} M.~L.,  2015, \mn@doi [\apjl] {10.1088/2041-8205/807/1/L12}, \href {https://ui.adsabs.harvard.edu/abs/2015ApJ...807L..12O} {807, L12}

\bibitem[\protect\citeauthoryear{{Ocvirk}, {Lewis}, {Gillet}, {Chardin}, {Aubert}, {Deparis}  \& {Th{\'e}lie}}{{Ocvirk} et~al.}{2021}]{Ocvirk.etal.2021}
{Ocvirk} P.,  {Lewis} J. S.~W.,  {Gillet} N.,  {Chardin} J.,  {Aubert} D.,  {Deparis} N.,   {Th{\'e}lie} {\'E}.,  2021, \mn@doi [\mnras] {10.1093/mnras/stab2502}, \href {https://ui.adsabs.harvard.edu/abs/2021MNRAS.507.6108O} {507, 6108}

\bibitem[\protect\citeauthoryear{{Pallottini} \& {Ferrara}}{{Pallottini} \& {Ferrara}}{2023}]{Pallottini.Ferrara.2023}
{Pallottini} A.,  {Ferrara} A.,  2023, \mn@doi [\aap] {10.1051/0004-6361/202347384}, \href {https://ui.adsabs.harvard.edu/abs/2023A&A...677L...4P} {677, L4}

\bibitem[\protect\citeauthoryear{{Pan} \& {Kravtsov}}{{Pan} \& {Kravtsov}}{2023}]{Pan.Kravtsov.2023}
{Pan} Y.,  {Kravtsov} A.,  2023, \mn@doi [arXiv e-prints] {10.48550/arXiv.2310.08636}, \href {https://ui.adsabs.harvard.edu/abs/2023arXiv231008636P} {p. arXiv:2310.08636}

\bibitem[\protect\citeauthoryear{{Pawlik}, {Schaye}  \& {van Scherpenzeel}}{{Pawlik} et~al.}{2009}]{Pawlik.etal.2009}
{Pawlik} A.~H.,  {Schaye} J.,   {van Scherpenzeel} E.,  2009, \mn@doi [\mnras] {10.1111/j.1365-2966.2009.14486.x}, \href {https://ui.adsabs.harvard.edu/abs/2009MNRAS.394.1812P} {394, 1812}

\bibitem[\protect\citeauthoryear{{P{\'e}rez-Gonz{\'a}lez} et~al.,}{{P{\'e}rez-Gonz{\'a}lez} et~al.}{2023}]{Perez.Gonzalez.etal.2024}
{P{\'e}rez-Gonz{\'a}lez} P.~G.,  et~al., 2023, \mn@doi [\apjl] {10.3847/2041-8213/acd9d0}, \href {https://ui.adsabs.harvard.edu/abs/2023ApJ...951L...1P} {951, L1}

\bibitem[\protect\citeauthoryear{{Planck Collaboration} et~al.,}{{Planck Collaboration} et~al.}{2016}]{Planck.coll.2016}
{Planck Collaboration} et~al., 2016, \mn@doi [\aap] {10.1051/0004-6361/201628897}, \href {https://ui.adsabs.harvard.edu/abs/2016A&A...596A.108P} {596, A108}

\bibitem[\protect\citeauthoryear{{Planck Collaboration} et~al.,}{{Planck Collaboration} et~al.}{2020}]{Planck.coll.2020}
{Planck Collaboration} et~al., 2020, \mn@doi [\aap] {10.1051/0004-6361/201833910}, \href {https://ui.adsabs.harvard.edu/abs/2020A&A...641A...6P} {641, A6}

\bibitem[\protect\citeauthoryear{{Qin}, {Mesinger}, {Bosman}  \& {Viel}}{{Qin} et~al.}{2021}]{Qin.etal.2021}
{Qin} Y.,  {Mesinger} A.,  {Bosman} S. E.~I.,   {Viel} M.,  2021, \mn@doi [\mnras] {10.1093/mnras/stab1833}, \href {https://ui.adsabs.harvard.edu/abs/2021MNRAS.506.2390Q} {506, 2390}

\bibitem[\protect\citeauthoryear{{Qin} et~al.,}{{Qin} et~al.}{2025}]{Qin.etal.2025}
{Qin} Y.,  et~al., 2025, \mn@doi [\pasa] {10.1017/pasa.2025.35}, \href {https://ui.adsabs.harvard.edu/abs/2025PASA...42...49Q} {42, e049}

\bibitem[\protect\citeauthoryear{{Robertson}}{{Robertson}}{2022}]{Robertson.2022}
{Robertson} B.~E.,  2022, \mn@doi [\araa] {10.1146/annurev-astro-120221-044656}, \href {https://ui.adsabs.harvard.edu/abs/2022ARA&A..60..121R} {60, 121}

\bibitem[\protect\citeauthoryear{{Robertson}, {Ellis}, {Furlanetto}  \& {Dunlop}}{{Robertson} et~al.}{2015}]{Robertson.etal.2015}
{Robertson} B.~E.,  {Ellis} R.~S.,  {Furlanetto} S.~R.,   {Dunlop} J.~S.,  2015, \mn@doi [\apjl] {10.1088/2041-8205/802/2/L19}, \href {https://ui.adsabs.harvard.edu/abs/2015ApJ...802L..19R} {802, L19}

\bibitem[\protect\citeauthoryear{{Robertson} et~al.,}{{Robertson} et~al.}{2024}]{Robertson.etal.2024}
{Robertson} B.,  et~al., 2024, \mn@doi [\apj] {10.3847/1538-4357/ad463d}, \href {https://ui.adsabs.harvard.edu/abs/2024ApJ...970...31R} {970, 31}

\bibitem[\protect\citeauthoryear{{Rosdahl} et~al.,}{{Rosdahl} et~al.}{2022}]{Rosdahl.etal.2022}
{Rosdahl} J.,  et~al., 2022, \mn@doi [\mnras] {10.1093/mnras/stac1942}, \href {https://ui.adsabs.harvard.edu/abs/2022MNRAS.515.2386R} {515, 2386}

\bibitem[\protect\citeauthoryear{{Saldana-Lopez} et~al.,}{{Saldana-Lopez} et~al.}{2023}]{Saldana_Lopez.etal.2023}
{Saldana-Lopez} A.,  et~al., 2023, \mn@doi [\mnras] {10.1093/mnras/stad1283}, \href {https://ui.adsabs.harvard.edu/abs/2023MNRAS.522.6295S} {522, 6295}

\bibitem[\protect\citeauthoryear{{Saxena} et~al.,}{{Saxena} et~al.}{2024}]{Saxena.etal.2024}
{Saxena} A.,  et~al., 2024, \mn@doi [\aap] {10.1051/0004-6361/202347132}, \href {https://ui.adsabs.harvard.edu/abs/2024A&A...684A..84S} {684, A84}

\bibitem[\protect\citeauthoryear{{Seiler}, {Hutter}, {Sinha}  \& {Croton}}{{Seiler} et~al.}{2019}]{Seiler.etal.2019}
{Seiler} J.,  {Hutter} A.,  {Sinha} M.,   {Croton} D.,  2019, \mn@doi [\mnras] {10.1093/mnras/stz1663}, \href {https://ui.adsabs.harvard.edu/abs/2019MNRAS.487.5739S} {487, 5739}

\bibitem[\protect\citeauthoryear{{Sharma}, {Theuns}, {Frenk}, {Bower}, {Crain}, {Schaller}  \& {Schaye}}{{Sharma} et~al.}{2016}]{Sharma.etal.2016}
{Sharma} M.,  {Theuns} T.,  {Frenk} C.,  {Bower} R.,  {Crain} R.,  {Schaller} M.,   {Schaye} J.,  2016, \mn@doi [\mnras] {10.1093/mnrasl/slw021}, \href {https://ui.adsabs.harvard.edu/abs/2016MNRAS.458L..94S} {458, L94}

\bibitem[\protect\citeauthoryear{{Sharma}, {Theuns}, {Frenk}, {Bower}, {Crain}, {Schaller}  \& {Schaye}}{{Sharma} et~al.}{2017}]{Sharma.etal.2017}
{Sharma} M.,  {Theuns} T.,  {Frenk} C.,  {Bower} R.~G.,  {Crain} R.~A.,  {Schaller} M.,   {Schaye} J.,  2017, \mn@doi [\mnras] {10.1093/mnras/stx578}, \href {https://ui.adsabs.harvard.edu/abs/2017MNRAS.468.2176S} {468, 2176}

\bibitem[\protect\citeauthoryear{{Shull}, {Harness}, {Trenti}  \& {Smith}}{{Shull} et~al.}{2012}]{Shull.etal.2012}
{Shull} J.~M.,  {Harness} A.,  {Trenti} M.,   {Smith} B.~D.,  2012, \mn@doi [\apj] {10.1088/0004-637X/747/2/100}, \href {https://ui.adsabs.harvard.edu/abs/2012ApJ...747..100S} {747, 100}

\bibitem[\protect\citeauthoryear{{Simmonds} et~al.,}{{Simmonds} et~al.}{2024a}]{Simmonds.etal.2024a}
{Simmonds} C.,  et~al., 2024a, \mn@doi [\mnras] {10.1093/mnras/stad3605}, \href {https://ui.adsabs.harvard.edu/abs/2024MNRAS.527.6139S} {527, 6139}

\bibitem[\protect\citeauthoryear{{Simmonds} et~al.,}{{Simmonds} et~al.}{2024b}]{Simmonds.etal.2024c}
{Simmonds} C.,  et~al., 2024b, \mn@doi [\mnras] {10.1093/mnras/stae2537}, \href {https://ui.adsabs.harvard.edu/abs/2024MNRAS.535.2998S} {535, 2998}

\bibitem[\protect\citeauthoryear{{Sokasian}, {Abel}  \& {Hernquist}}{{Sokasian} et~al.}{2003}]{Sokasian.etal.2003}
{Sokasian} A.,  {Abel} T.,   {Hernquist} L.,  2003, \mn@doi [\mnras] {10.1046/j.1365-8711.2003.06308.x}, \href {https://ui.adsabs.harvard.edu/abs/2003MNRAS.340..473S} {340, 473}

\bibitem[\protect\citeauthoryear{{Spina}, {Bosman}, {Davies}, {Gaikwad}  \& {Zhu}}{{Spina} et~al.}{2024}]{Spina.etal.2024}
{Spina} B.,  {Bosman} S. E.~I.,  {Davies} F.~B.,  {Gaikwad} P.,   {Zhu} Y.,  2024, \mn@doi [\aap] {10.1051/0004-6361/202450798}, \href {https://ui.adsabs.harvard.edu/abs/2024A&A...688L..26S} {688, L26}

\bibitem[\protect\citeauthoryear{{Sun}, {Faucher-Gigu{\`e}re}, {Hayward}, {Shen}, {Wetzel}  \& {Cochrane}}{{Sun} et~al.}{2023}]{Sun.etal.2023}
{Sun} G.,  {Faucher-Gigu{\`e}re} C.-A.,  {Hayward} C.~C.,  {Shen} X.,  {Wetzel} A.,   {Cochrane} R.~K.,  2023, \mn@doi [\apjl] {10.3847/2041-8213/acf85a}, \href {https://ui.adsabs.harvard.edu/abs/2023ApJ...955L..35S} {955, L35}

\bibitem[\protect\citeauthoryear{{Tacchella}, {Forbes}  \& {Caplar}}{{Tacchella} et~al.}{2020}]{Tacchella.etal.2020}
{Tacchella} S.,  {Forbes} J.~C.,   {Caplar} N.,  2020, \mn@doi [\mnras] {10.1093/mnras/staa1838}, \href {https://ui.adsabs.harvard.edu/abs/2020MNRAS.497..698T} {497, 698}

\bibitem[\protect\citeauthoryear{{Tinker}, {Kravtsov}, {Klypin}, {Abazajian}, {Warren}, {Yepes}, {Gottl{\"o}ber}  \& {Holz}}{{Tinker} et~al.}{2008}]{Tinker.etal.2008}
{Tinker} J.,  {Kravtsov} A.~V.,  {Klypin} A.,  {Abazajian} K.,  {Warren} M.,  {Yepes} G.,  {Gottl{\"o}ber} S.,   {Holz} D.~E.,  2008, \mn@doi [\apj] {10.1086/591439}, \href {https://ui.adsabs.harvard.edu/abs/2008ApJ...688..709T} {688, 709}

\bibitem[\protect\citeauthoryear{{Topping}, {Stark}, {Endsley}, {Plat}, {Whitler}, {Chen}  \& {Charlot}}{{Topping} et~al.}{2022}]{Topping.etal.2022}
{Topping} M.~W.,  {Stark} D.~P.,  {Endsley} R.,  {Plat} A.,  {Whitler} L.,  {Chen} Z.,   {Charlot} S.,  2022, \mn@doi [\apj] {10.3847/1538-4357/aca522}, \href {https://ui.adsabs.harvard.edu/abs/2022ApJ...941..153T} {941, 153}

\bibitem[\protect\citeauthoryear{{Weibel} et~al.,}{{Weibel} et~al.}{2026}]{Weibel.etal.2026}
{Weibel} A.,  et~al., 2026, \mn@doi [\apj] {10.3847/1538-4357/ae5a9c}, \href {https://ui.adsabs.harvard.edu/abs/2026ApJ..1002..136W} {1002, 136}

\bibitem[\protect\citeauthoryear{{Weisz} \& {Boylan-Kolchin}}{{Weisz} \& {Boylan-Kolchin}}{2017}]{Weisz.BoylanKolchin.2017}
{Weisz} D.~R.,  {Boylan-Kolchin} M.,  2017, \mn@doi [\mnras] {10.1093/mnrasl/slx043}, \href {https://ui.adsabs.harvard.edu/abs/2017MNRAS.469L..83W} {469, L83}

\bibitem[\protect\citeauthoryear{{Whitler} et~al.,}{{Whitler} et~al.}{2025}]{Whitler.etal.2025}
{Whitler} L.,  et~al., 2025, \mn@doi [\apj] {10.3847/1538-4357/adfddc}, \href {https://ui.adsabs.harvard.edu/abs/2025ApJ...992...63W} {992, 63}

\bibitem[\protect\citeauthoryear{{Wise}, {Demchenko}, {Halicek}, {Norman}, {Turk}, {Abel}  \& {Smith}}{{Wise} et~al.}{2014}]{Wise.etal.2014}
{Wise} J.~H.,  {Demchenko} V.~G.,  {Halicek} M.~T.,  {Norman} M.~L.,  {Turk} M.~J.,  {Abel} T.,   {Smith} B.~D.,  2014, \mn@doi [\mnras] {10.1093/mnras/stu979}, \href {https://ui.adsabs.harvard.edu/abs/2014MNRAS.442.2560W} {442, 2560}

\bibitem[\protect\citeauthoryear{{Wu} \& {Kravtsov}}{{Wu} \& {Kravtsov}}{2024}]{Wu.Kravtsov.2024}
{Wu} Z.,  {Kravtsov} A.,  2024, \mn@doi [The Open Journal of Astrophysics] {10.33232/001c.121193}, \href {https://ui.adsabs.harvard.edu/abs/2024OJAp....7E..56W} {7, 56}

\bibitem[\protect\citeauthoryear{{Yue}, {Ferrara}  \& {Xu}}{{Yue} et~al.}{2016}]{Yue.etal.2016}
{Yue} B.,  {Ferrara} A.,   {Xu} Y.,  2016, \mn@doi [\mnras] {10.1093/mnras/stw2145}, \href {https://ui.adsabs.harvard.edu/abs/2016MNRAS.463.1968Y} {463, 1968}

\bibitem[\protect\citeauthoryear{{Zhu} et~al.,}{{Zhu} et~al.}{2022}]{Zhu.etal.2022}
{Zhu} Y.,  et~al., 2022, \mn@doi [\apj] {10.3847/1538-4357/ac6e60}, \href {https://ui.adsabs.harvard.edu/abs/2022ApJ...932...76Z} {932, 76}

\bibitem[\protect\citeauthoryear{{Zhu} et~al.,}{{Zhu} et~al.}{2024}]{Zhu.etal.2024}
{Zhu} Y.,  et~al., 2024, \mn@doi [\mnras] {10.1093/mnrasl/slae061}, \href {https://ui.adsabs.harvard.edu/abs/2024MNRAS.533L..49Z} {533, L49}

\bibitem[\protect\citeauthoryear{{{\v{D}}urov{\v{c}}{\'\i}kov{\'a}}, {Katz}, {Bosman}, {Davies}, {Devriendt}  \& {Slyz}}{{{\v{D}}urov{\v{c}}{\'\i}kov{\'a}} et~al.}{2020}]{Durovcikova.etal.2020}
{{\v{D}}urov{\v{c}}{\'\i}kov{\'a}} D.,  {Katz} H.,  {Bosman} S. E.~I.,  {Davies} F.~B.,  {Devriendt} J.,   {Slyz} A.,  2020, \mn@doi [\mnras] {10.1093/mnras/staa505}, \href {https://ui.adsabs.harvard.edu/abs/2020MNRAS.493.4256D} {493, 4256}

\bibitem[\protect\citeauthoryear{{{\v{D}}urov{\v{c}}{\'\i}kov{\'a}} et~al.,}{{{\v{D}}urov{\v{c}}{\'\i}kov{\'a}} et~al.}{2024}]{Durovcikova.etal.2024}
{{\v{D}}urov{\v{c}}{\'\i}kov{\'a}} D.,  et~al., 2024, \mn@doi [\apj] {10.3847/1538-4357/ad4888}, \href {https://ui.adsabs.harvard.edu/abs/2024ApJ...969..162D} {969, 162}

\makeatother
\end{thebibliography}

\hfill \break
\hfill \break
\hfill \break

\appendix

\section{$\fesc$ and $C_R$ models}
\label{app:model_plots}

In this section, we include visualizations of the escape fraction $\fesc$ and the IGM clumping factor $C_R$ models we use in this work.

\begin{figure}
   \centering {
   \includegraphics[width=0.39\textwidth]{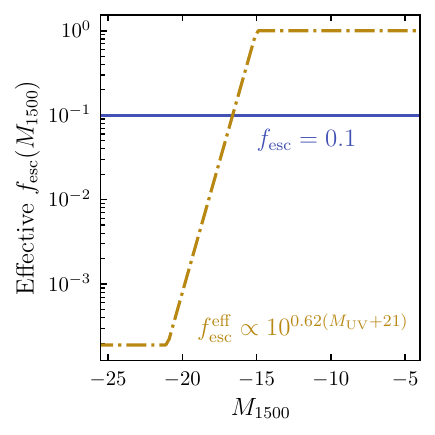}
   }
   \caption{
        Two of the three $\fesc$ models we adopt with respect to galaxy luminosity $\Muv$. For details, see \S~\ref{subsec:fesc_models}.
    }
   \label{fig:fesc_z}
\end{figure}

Figure~\ref{fig:fesc_z} shows the two of the three $\fesc$ models we adopt with respect to galaxy luminosity $\Muv$: a constant global $\fesc = 0.1$ (blue), and a model in which $\fesc = \fesc(\Muv = -21) \times 10 ^ {0.62(\Muv + 21)}$ increases monotonically for fainter galaxies (yellow), where $\fesc(\Muv = -21) = 1.91\times 10^{-4}$. The third model where $\fesc$ depends on specific star formation rate cannot be plotted with $\Muv$, and is presented in Figure~\ref{fig:SPHINX_sSFR} instead. For details, see Section~\ref{subsec:fesc_models}.

\begin{figure}
   \centering {
   \includegraphics[width=0.39\textwidth]{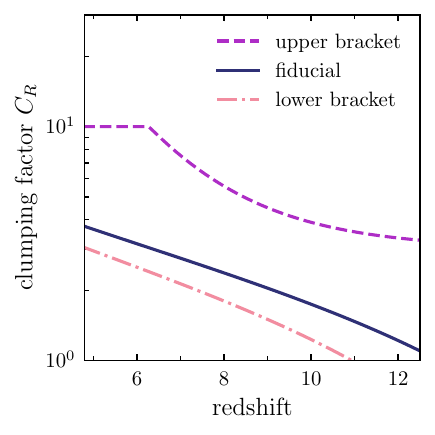}
   }
   \caption{
        The IGM clumping factor $C_R$ in three different $C_R(z)$ models considered in our study (see \S~\ref{subsec:hydrogen_ion_modeling} and \ref{subsec:clumping_factor} for details).
    }
   \label{fig:clump_z}
\end{figure}

Figure~\ref{fig:clump_z} shows the three models $C_R(z)$ considered in this study.
For details on our choices of models, see Section~\ref{subsec:hydrogen_ion_modeling}. The effect of different $C_R$ assumptions is discussed in Section~\ref{subsec:clumping_factor}.

\section{Stochastic formation parameters}
\label{app:stochastic_params}

Our star formation model used \texttt{GRUMPY} with specific modifications for modeling reionization as detailed in \citealt{Wu.Kravtsov.2024}, which has found a good match with $5 < z < 10$ galaxy properties like the UV luminosity function. However, to account for the increasing effect of feedback in high redshift $z > 10$ galaxies, we modify our prescription of bursty star formation in line with \citet{Kravtsov.Belokurov.2024}, which adopted a SFR stochasticity $\sigma_\Delta$ increasing with redshift.
We follow \citet{Caplar.Tacchella.2019} and use the PSD of the form ${\rm PSD}(f) = \sigma^2_\Delta[1+(\tau_{\rm break}f)^\alpha]^{-1}$,
where $\sigma_\Delta$ characterizes the amplitude of the SFR variability over long time scales and  $\tau_{\rm break}$ characterizes the timescale over which the random numbers are effectively uncorrelated. Parameter $\alpha$ controls the slope of the PSD at high frequencies (short time scales). 
In our models, we fix the slope $\alpha$ and $\tau_{\rm break}$ to the values $\alpha=2$ and $\tau_{\rm break}=100$ Myr, which are physically motivated by the time scales of gas evolution and star formation in giant molecular clouds in a typical ISM \citep[see][for a detailed discussion]{Tacchella.etal.2020}. For $z = 5-10$, we use $\sigma_\Delta=0.1$ consistent with the typical amount of SFR stochasticity in host halos of observed galaxies at these redshifts. While for $z > 10$ we take $\sigma_\Delta$ values consistent with \citet{Kravtsov.Belokurov.2024}. The specific parameters per redshift are summarized in Table~\ref{tab:stoch_params}.

\begin{table}
    \centering
	\caption{
    Stochasticity parameters used for galaxy models of different redshifts $z = 5 - 16$, consistent with \citet{Kravtsov.Belokurov.2024}.
 }
	\begin{tabular}{ccccccc}
		\hline\hline\\[-2mm]
		$z$ & $5 - 10$ & 11 & 12 & 13 & 14 & 16 \\[1mm]
		\hline\\[-2mm]
            $z_{\rm init}$ & 25 & 35 & 35 & 35 & 35 & 35 \\
            $\sigma_\Delta$ & 0.1 & 0.15 & 0.15 & 0.16 & 0.18 & 0.25 \\[1mm]
            $\sigma_{M_{\rm UV}}$ & 0.7 & 1.1 & 1.4 & 1.5 & 1.7 & 2.2 \\[1mm]
    \hline
    \label{tab:stoch_params}
    \end{tabular}
\end{table}

The corresponding scatter in the UV absolute magnitude is $\sigma_{\rm M_{\rm UV}}\approx 0.75$ at low redshifts, increasing to $\sim 2$ at $z\approx16$. These values are broadly consistent with the UV absolute magnitude fluctuations estimated in high-resolution zoom-in cosmological simulations at the same redshifts \citep[with slightly different values]{Pallottini.Ferrara.2023,Sun.etal.2023}. We additionally note that adding stochasticity in our high-redshift galaxy model does not impact the late-time agreement with local universe observations. \citep[see][for further exploration of the effects of stochasticity in the context of the model we use]{Kravtsov.Belokurov.2024}

\section{UV LF modified Schechter fits}
\label{app:uvlf_fit}

\begin{table}
    \centering
	\caption{
 Best-fit parameters for \citet{Jaacks.etal.2013}'s modified Schechter function to our stochastic UV LF at $z = 5 - 16$. The last column $\alpha - \beta$ shows the effective faint end LF slope.
 }
	\begin{tabular}{ccccccc}
		\hline\hline\\[-2mm]
		$z$ & $\log_{10}\phis$ & $M_{1500, *}$ & $M_{1500, t}$ & $\alpha$ & $\beta$ & $\alpha-\beta$ \\
            & ${\rm Mpc^{-3}}$ & & & & & \\[1mm]
		\hline\\[-2mm]
            5 & -2.640 & -23.268 & -17.513 & -0.453 & 0.744 & -1.197 \\
            6 & -3.544 & -23.736 & -18.668 & -0.665 & 0.792 & -1.458 \\
            7 & -3.695 & -23.502 & -17.653 & -0.702 & 0.789 & -1.491 \\
            8 & -4.246 & -23.846 & -17.957 & -0.772 & 0.986 & -1.758 \\
            9 & -3.979 & -22.604 & -16.761 & -0.786 & 0.793 & -1.579 \\
            10 & -4.428 & -22.448 & -16.812 & -0.863 & 0.863 & -1.726 \\
            11 & -3.681 & -22.546 & -13.994 & -0.756 & 0.577 & -1.333 \\
            12 & -3.611 & -22.232 & -12.348 & -0.748 & 0.598 & -1.346 \\
            13 & -3.173 & -22.387 & -11.161 & -0.661 & 0.655 & -1.316 \\
            14 & -2.393 & -22.056 & -10.012 & -0.544 & 0.713 & -1.257 \\
            16 & -0.669 & -22.229 & -19.979 & -0.229 & 0.926 & -1.155 \\[1mm]
    \hline
    \label{tab:jaacks_params}
    \end{tabular}
\end{table}

We approximate model UV luminosity functions with the modified Schechter functional form of \citet{Jaacks.etal.2013}:
\begin{equation}
    \Phi(L) = \phis {\left(\frac{L}{\Ls}\right)}^{\alpha} \exp{\left(-\frac{L}{\Ls}\right)} {\left[1+{\left(\frac{L}{\Lt}\right)}^{\beta}\right]}^{-1},
    \label{eq:jaacks}
\end{equation}
where $\phis$ and $\Ls$ are the normalization and characteristic luminosity of the bright end, respectively.
Compared to the Schechter form, which has a fixed faint-end slope $\alpha$, this form has a slope that can become progressively shallower or steeper around $\Lt$ and reaches the asymptotic slope of $\alpha-\beta$ at $L\ll \Lt$.

We determine the best-fit parameters of the function by minimizing the least-squares differences between the functional form and model UV LF converted from the luminosity to absolute magnitude $\Muv$ using the conversion
\begin{eqnarray}
\Muv &=& -2.5\log_{10} \frac{\Luv}{4\pi (10\,{\rm pc})^2} - 48.6\nonumber\\ &=& -2.5\log_{10} \Luv + 51.59,
\end{eqnarray}
where $L_{\rm UV }$ is the luminosity density at $\lambda=1500\,\aa$ in $\rm egs\,s^{-1}\, Hz^{-1}$. 

The best-fit parameters for different redshifts $5 < z < 16$ are presented in Table~\ref{tab:jaacks_params}. The table also shows the effect of reionization on the UV LF: it shows significant flattening of the faint-end slope $\alpha-\beta$ for redshifts $z \lesssim 6$ due to suppression of accretion caused by the UV heating of the intergalactic medium during and after reionization. In what follows, we present analytical fits to the emission rate of ionizing photons by UV magnitude, using the same modified Schechter form.

\section{$\nion$ functional fit parameter}
\label{app:nion_params}

\begin{figure*}
   \centering {
   \includegraphics[width=0.99\textwidth]{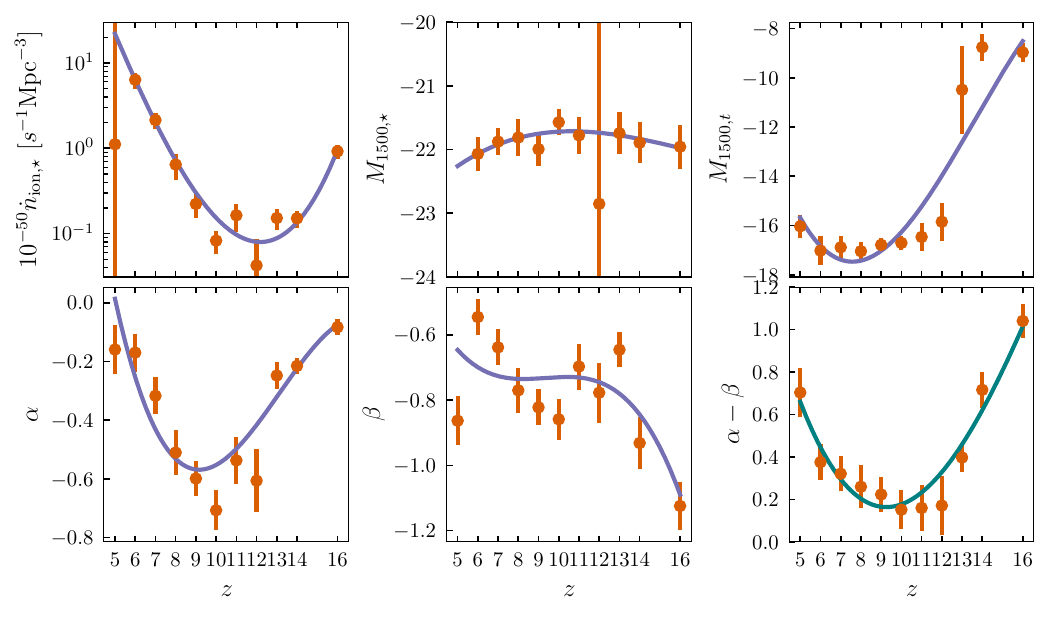}
   }
   \caption{
   Coefficients of the third-order polynomial fit approximation to the evolution of the parameters of the \citet{Jaacks.etal.2013} analytical form to $\dnion(\Muv)$ at $5\leq z\leq 16$. The bottom right panel (green line) shows how the faint-end slope of the $\dnion$ functions change across redshifts $5 < z < 16$. The polynomial coefficients are presented in Table~\ref{tab:nion_poly_fit}.
   }
   \label{fig:nion_polynomial}
\end{figure*}

We construct an ionizing flux function as a function of $\Muv$, $\dot{n}_{\rm ion}(\Muv)$ similarly to how we estimate model UV LF (Section~\ref{subsec:comp_uv_ion}). Namely, we compute $\dot{n}_{\rm ion}(\Muv)$ as a weighted histogram of halos in a box of a given comoving size $V_{\rm box}$ in bins of $\Muv$ with weights given by $\dot{N}_{\rm ion}/(f V_{\rm box})$, where $\dot{N}_{\rm ion}$ is the Lyman continuum photom emission rate of each model galaxy and $f=f(\M200c)$ is the fraction of selected halos in a given box.

This function can also be approximated analytically by the same modified Schechter function from \citet{Jaacks.etal.2013}, in which  $\phis$ in Equation~\ref{eq:jaacks} is replaced with ${\dot n}_{\rm ion, *}$. Fit parameters for the ionizing photon flux are shown in Table~\ref{tab:nion_params}, analogous to $M_{1500}$ UV LF parameters presented earlier in Table~\ref{tab:jaacks_params}.

\begin{table}
    \centering
	\caption{
 Best-fit parameters for \citet{Jaacks.etal.2013}'s modified Schechter function to the ionizing flux function $\dot{n}_{\rm ion}$ at $5\leq z\leq 16$.
 }
	\begin{tabular}{cccccc}
		\hline\hline\\[-2mm]
		$z$ & $\log_{10} 10^{-50}{\dot n}_{\rm ion, *}$ & $M_{1500, *}$ & $M_{1500, t}$ & $\alpha$ & $\beta$ \\[1mm]
		\hline\\[-2mm]
            5 & -3.914 & -22.375 & -15.184 & -0.007 & -0.738\\
            6 & -3.197 & -22.067 & -17.017 & -0.170 & -0.546\\
            7 & -2.670 & -21.875 & -16.883 & -0.317 & -0.639\\
            8 & -2.190 & -21.810 & -17.042 & -0.510 & -0.770\\
            9 & -0.652 & -21.993 & -16.783 & -0.599 & -0.822\\
            10 & -1.082 & -21.571 & -16.704 & -0.707 & -0.859\\
            11 & -0.786 & -21.776 & -16.462 & -0.537 & -0.697\\
            12 & -1.372 & -22.852 & -15.845 & -0.606 & -0.777\\
            13 & -0.818 & -21.742 & -10.486 & -0.248 & -0.646\\
            14 & -0.821 & -21.892 & -8.758 & -0.215 & -0.932\\
            16 & -0.036 & -21.955 & -8.966 & -0.083 & -1.124\\[1mm]
    \hline
    \label{tab:nion_params}
    \end{tabular}
\end{table}

We provide an approximation for how best-fit parameters of the functional form evolve with redshift for $5\leq z\leq 16$, so that one can reproduce our calculations of the LyC photon budget, and model the ionization history of the Universe by solving the time-evolution of $\dot{n}_{\rm ion}$ using Equation~\ref{eq:dqdz}. Namely, we approximate the evolution of each parameter using third-order polynomials ${\rm param} = a_0 + a_1 z + a_2 z^2 + a_3 z^3$. Figure~\ref{fig:nion_polynomial} shows the polynomial fits for each parameter, while Table~\ref{tab:nion_poly_fit} presents the best-fit values of $a_i$ coefficients.\\[3mm]

\begin{table}
    \centering
	\caption{
 Coefficients of the third-order polynomial fit approximation to the evolution of the parameters of the \citet{Jaacks.etal.2013} approximation to $\dot{n}_{\rm ion}(\Muv)$ with redshift at $5\leq z\leq 16$.
 }
	\begin{tabular}{ccccc}
		\hline\hline\\[-2mm]
         Parameters & $a_3$ & $a_2$ & $a_1$ & $a_0$ \\[1mm]
         $10^{-50}{\dot n}_{\rm ion, *}$ & 0.0024 & -0.0242 & -0.4948 & 4.1152 \\
         $M_{1500, *}$ & 0.0007 & -0.0367 & 0.5284 & -24.0759 \\
         $M_{1500, t}$ & -0.0130 & 0.5310 & -5.8082 & 1.7121 \\
         $\alpha$ & -0.0021 & 0.0824 & -0.9830 & 3.1297 \\
         $\beta$ & -0.0013 & 0.0365 & -0.3356 & 0.2826 \\[1mm]
    \hline
    \label{tab:nion_poly_fit}
    \end{tabular}
\end{table}


\end{document}